\documentclass[useAMS,usegraphicx]{mn2e}

\def\simlt{\mathrel{\hbox{\rlap{\hbox{\lower4pt\hbox{$\sim$}}}\hbox{$<$}}}}
\def\simgt{\mathrel{\hbox{\rlap{\hbox{\lower4pt\hbox{$\sim$}}}\hbox{$>$}}}}

\def\ale{\mathrel{\hbox{\rlap{\hbox{\lower4pt\hbox{$\sim$}}}\hbox{$<$}}}}
\def\age{\mathrel{\hbox{\rlap{\hbox{\lower4pt\hbox{$\sim$}}}\hbox{$>$}}}}

\title[Nearby Supernova Rates from LOSS: I. The Methods and Database]{
Nearby Supernova Rates from the Lick Observatory Supernova Search.
I. The Methods and Database}

\author[Leaman et al.]{Jesse Leaman$^{1,2}$\thanks{E-mail: jleaman@astro.berkeley.edu (JL),
wli@astro.berkeley.edu (WL), 
rchornock@cfa.harvard.edu (RC), 
alex@astro.berkeley.edu (AVF)}, 
Weidong Li$^{1}$\footnotemark[1], 
Ryan Chornock$^{1,3}$\footnotemark[1], and Alexei V. Filippenko$^{1}$\footnotemark[1] \\ 
$^{1}$Department of Astronomy, University of California,
Berkeley, CA 94720-3411, USA\\
$^{2}$NASA Ames Research Center, Mountain View, CA 94043, USA\\
$^{3}$Harvard-Smithsonian Center for Astrophysics, 60 Garden Street, 
Cambridge, MA 02138, USA}

\begin{document}
\maketitle

\begin{abstract}

This is the first paper of a series in which we present new
measurements of the observed rates of supernovae (SNe) in the local
Universe, determined from the Lick Observatory Supernova Search
(LOSS). We have obtained 2.3 million observations of 14,882 sample
galaxies over an interval of 11 years (March 1998 through
Dec. 2008). We considered 1036 SNe detected in our sample and used an
optimal subsample of 726 SNe (274 SNe~Ia, 116 SNe~Ibc, 324 SNe~II) to
determine our SN rates. This is the largest and most homogeneous set
of nearby SNe ever assembled for this purpose, and ours is the first
local SN rate analysis based on CCD imaging and modern
image-subtraction techniques.  In this paper, we lay the foundation of
the study. We derive the recipe for the control-time calculation for
SNe with a known luminosity function, and provide details on the
construction of the galaxy and SN samples used in the
calculations. Compared with a complete volume-limited galaxy sample,
our sample has a deficit of low-luminosity galaxies but still provides
enough statistics for a reliable rate calculation. There is a strong
Malmquist bias, so the average size (luminosity or mass) of the
galaxies increases monotonically with distance, and this trend is used
to showcase a correlation between SN rates and galaxy sizes. Very few
core-collapse SNe are found in early-type galaxies, providing strong
constraints on the amount of recent star formation within these
galaxies.  The small average observation interval ($\sim 9$~d) of
our survey ensures that our control-time calculations can tolerate a
reasonable amount of uncertainty in the luminosity functions of
SNe. We perform Monte Carlo simulations to determine the limiting
magnitude of each image and the SN detection efficiency as a function
of galaxy Hubble type.  The limiting magnitude and the detection
efficiency, together with the luminosity function derived from a
complete sample of very nearby SNe in Paper II, will be used to
calculate the control time and the SN rates in Paper III.

\end{abstract}


\section{Introduction}

Supernovae (SNe), which occur in several spectroscopically distinct
varieties (e.g., Wheeler \& Harkness 1990; Filippenko 1997), represent
the final, explosive stage in the evolution of certain varieties of
stars, and are among the most interesting and important constituents
of the Universe. SNe provide a celestial laboratory to study stellar
evolution.  They synthesise and expel heavy elements, thereby
dictating much of the chemical evolution of galaxies. Shock waves from
SNe inject energy into the interstellar media of galaxies and may also
trigger vigorous bursts of star formation.  They lead to 
the formation of neutron stars, and probably even black holes under
some circumstances. Some SNe are associated with energetic gamma-ray
bursts.  Being so powerful and calibratable in their observed
properties, Type Ia supernovae (SNe~Ia) are exceedingly useful
cosmological probes; they led to the discovery of the accelerating
expansion of the Universe (Riess et al. 1998; Perlmutter et al. 1999;
see Filippenko 2005 for a review).

The rate of SNe is a key quantity for astrophysics. Measurements of
the SN rate and its evolution over cosmic time provide important
information on the metal enrichment and chemical evolution of the
Universe; the structure, kinematics, and composition of the
interstellar medium; the birth rate of compact objects; and the
formation of SNe from different types of progenitor systems.

Over the past seventy years, many studies have attempted to measure
the SN rate in the local Universe and at moderate to high
redshifts. There are both advantages and disadvantages of measuring
the SN rate in nearby galaxies.  The nearby SNe are bright and can be
easily observed with small to moderate-sized telescopes. Information
on the host galaxies such as the Hubble type, luminosity, and colour
are more readily available for the nearby galaxies than for the more
distant ones, making it possible to do detailed studies of the rate
dependence on SN environment. On the other hand, nearby SNe are
relatively rare, so it takes a long time to accumulate enough
statistics.  It is also difficult to do an all-sky survey for the
nearby SNe; thus, most nearby searches have, in the past, conducted
targeted snapshots of a sample of known galaxies.  Determination of
the SN rate from such targeted surveys requires the ``control-time
method'' (see \S 3 and the Appendix), which historically has been
plagued with uncertainties in the photometric behaviour of SNe, their
luminosity function (LF), and the extinction they experience in their
host galaxies. The monitoring of a predetermined galaxy sample also
has the potential to introduce selection biases.

Van den Bergh (1991) summarises estimates of nearby SN rates before
the 1990s. These earlier rates are based mostly on the Palomar SN
search (Zwicky 1938, 1942), the Asiago SN search (Cappellaro \&
Turatto 1988), and Robert Evans' visual search (van den Bergh,
McClure, \& Evans 1987; van den Bergh \& McClure 1990). The SN rate
was first expressed as the frequency of SNe in an average galaxy, but
it was realised that this frequency is proportional to the host-galaxy
luminosity (Pskovskii 1961, 1967; however, see the discussion in Paper
III for a nonlinear proportionality). Accordingly, the SN rate is
generally normalised by the galaxy luminosity and given in units of
SNu, or 1 SN (100~yr 10$^{10}~ {\rm L}_\odot$)$^{-1}$.

In the past two decades, the most influential studies of nearby SN
rates were conducted by Cappellaro et al. (1993a, 1993b, 1997,
1999). In particular, Cappellaro et al.  (1999; C99, hereafter)
combined five surveys to increase the total number of SN discoveries
to 136.\footnote{As noted by Mannucci et al. (2005), one of the 137 SNe
  used in C99 was later discovered to be associated with a galaxy
  not included in the sample.}  The SN rates, normalised to the
$B$-band luminosity of the host galaxies, were measured for galaxies
of different Hubble types, and were compared to different tracers of
star formation (e.g., broad-band colours, far-infrared luminosities).

The C99 database is the heart and soul of numerous studies of the SN
rate. Mannucci et al. (2005; hereafter M05) normalised the rates with
the infrared $K$-band luminosity, as well as with the mass derived
from the $K$-band luminosity and $B - K$ colours of the galaxies.
Della Valle et al. (2005) found that radio-loud early-type galaxies
have a SN~Ia rate that is a factor of 4 higher than that of the
radio-quiet early-type galaxies, and suggested that the enhancement is
probably caused by repeated episodes of interaction and/or merger
events.  Mannucci et al. (2008) investigated the cluster early-type
galaxy SN~Ia rate, and found that it is more than three times larger
than that in field early-type galaxies, perhaps due to galaxy
interactions in the clusters.  Mannucci et al. (2005, 2006) further
derived delay-time distributions of SNe~Ia, and postulated that there
may be two components in the SN~Ia population: a ``prompt" component
with the SNe~Ia exploding soon after their stellar birth, and a
``tardy" component with the SNe~Ia exploding after a long delay
following star formation. The implications of the two-component model
on the progenitors of SNe~Ia and the impact on their use as
calibratable candles to derive cosmological parameters are further
discussed by Scannapieco \& Bildsten (2005), Sullivan et al. (2006),
and Dahlen et al. (2008).

Because of the growing interest in using the evolution of the SN~Ia
rate over cosmic time to constrain the progenitor systems of SNe~Ia,
numerous recent studies have derived SN~Ia rates at moderate to high
redshifts (e.g., Hardin et al. 2000; Madgwick et al.  2003; Tonry et
al. 2003; Blanc et al. 2004; Strolger et al. 2004; Dahlen et al. 2004,
2008; Neill et al. 2006; Poznanski et al. 2007; Botticella et
al. 2008; Dilday et al. 2008; Horesh et al. 2008). Many of these
investigations were conducted with systematic rolling searches using
large ground-based telescopes or the {\it Hubble Space Telescope},
yielding rates that have precisions comparable or superior to those of
the published nearby SN rates. Thus, it is critical to improve the
precision of the nearby SN rates before they become the bottleneck for
studies of the cosmic evolution of SN rates.

In this series of papers, we report the determination of nearby SN
rates from our long-term efforts with the Lick Observatory Supernova
Search (LOSS).  LOSS, conducted with the 0.76-m Katzman Automatic
Imaging Telescope (KAIT), has been described by Li et al. (2000),
Filippenko et al. (2001), Filippenko (2003, 2005), and Filippenko, Li,
\& Treffers (2011). KAIT, based on the prototype Berkeley
Automatic Imaging Telescope (Richmond, Treffers, \& Filippenko 1993),
is a fully robotic telescope whose control system checks the weather
and performs observations with a dedicated CCD camera without human
intervention. The data are automatically processed through an
image-subtraction pipeline (Ganeshalingam et al.  2010; Filippenko et
al. 2011), and candidate SNe are flagged. The next day, these
candidates are visually inspected by a group of students (primarily
undergraduate) in the Department of Astronomy at the University of
California, Berkeley. The most promising SN candidates are reobserved
the next clear night, and the confirmed SNe are reported to the
Central Bureau of Astronomical Telegrams (CBAT). During the period
from March 1998 through the end of 2008 (on which the data from this
study are based), LOSS found 732 SNe, easily exceeding any other
searches for nearby SNe and accounting for more than 40\% of all SNe
with redshift $z < 0.05$ reported to the Central Bureau for
Astronomical Telegrams.

This is Paper I of a series of several papers, and is organised as
follows. Section 2 provides an outline of the series, summarises the
improvements of our rate determination over the published results, and
discusses possible limitations of our study.  In \S 3 we describe the
control-time method for a family of light curves with a known
LF. Section 4 discusses the galaxy and SN samples, and other key
ingredients in the rate calculations, including the log files, the
limiting-magnitude determination, and the detection efficiency. Our
conclusions are summarised in \S 5. Throughout the study, we adopt the
WMAP5 Hubble constant of $H_0 = 73$ km s$^{-1}$ Mpc$^{-1}$ (Spergel et
al. 2007), consistent with the recent direct determination based on
Cepheid variables and SNe~Ia by Riess et al. (2009).

\section{Series Outline and Summary of Improvements} 

\subsection{Outline of the Series}

Paper I (this paper) lays the foundation for our study. It discusses
the control-time method for SNe with a known LF, and provides detailed
information on the galaxy and SN samples. The physical parameters
(Hubble type, colours, mass, inclination, etc.) of the LOSS sample
galaxies are studied to check whether they are typical of the normal
volume-limited galaxy distribution. The various subsamples of galaxies
and SNe are also described. The paper then discusses the log files
used in the rate calculations, and shows some statistics on the
observation-interval distribution in our search. We also report how
the limiting magnitudes are derived for each KAIT image based on
several parameters in the log files.  Finally, we use Monte Carlo
simulations to study the detection efficiency --- that is, how our
image-processing software handles detections of SNe with different
significance. The limiting magnitude and the detection efficiency are
key ingredients in the control-time calculations.

Paper II (Li et al. 2011a) reports the determination of the observed
LFs of SNe, an important piece of information for the control-time
calculation.  This is the first time such LFs have been derived; they
eliminate uncertainties in the historical SN rate calculations caused
by uncertainties in the assumed SN light curves, the peak-magnitude
distribution, and the host-galaxy extinction. We first constructed a
distance-limited sample of 175 SNe ($D < 60$~Mpc for the core-collapse
SNe~Ibc and II, $D < 80$~Mpc for the SNe~Ia), and then measured the
light curve for every SN from either the SN search images or the
dedicated follow-up photometry. A family of light curves was then used
to fit the peak observed magnitude of each SN, and the observed
absolute magnitude of the SN was calculated. For each SN, the
completeness of our survey is determined from the monitoring history.
We study the dependence of the SN LF on the host galaxy and SN subtype
properties, and report the observed fraction of different subtypes of
SNe and LFs in a magnitude-limited survey.

Paper III (Li et al. 2011b) uses all of the information presented in
the first two papers and calculates the control time for the KAIT
sample galaxies.  We show that our SN rates have a significant
correlation with the distance bin used for the calculation, and
demonstrate that this is caused by a ``rate-size relation'': the SN
rates, after being normalised linearly by the size (luminosity or
mass) of their host galaxies, are still correlated with the galaxy
size, in the sense that smaller galaxies have higher rates. A
rate-size slope (RSS, hereafter) is used to normalise the rates to the
same galaxy size. The RSSs for different normalisations are estimated
and are used to derive the rates.  The SN rates for different types of
SNe are reported for galaxies of different Hubble types and colours.
We compare our rates to the published C99/M05 results and find that
our rates, with more bins and smaller error bars, are in good
agreement with the published measurements when the rates are
calculated in the same manner. We also derive the expected SN rate in
the Milky Way Galaxy and the volumetric rates in the local
Universe. Discussions of the possible causes of the rate-size
relation, and of the two-component model fit for SN~Ia rates, are also
provided.

Using a subset of LOSS sample galaxies having Sloan Digital Sky Survey
(SDSS) spectra, Paper IV (Maoz et al. 2011) developed a robust method
to recover the SN delay time distribution (DTD) --- the SN rate versus
time that would follow a brief burst of star formation --- and found
strong evidence for populations of SNe~Ia in both ``prompt" (age $<
420$ Myr) and ``delayed" (age $> 2.4$ Gyr) stellar populations.

\subsection{Summary of Improvements over Published Results}

Our study improves upon previously published results in a number of
ways, as follows.

(1) {\it Increased sample size.} Small-number statistics have been a
key issue in calculations of nearby SN rates. The C99 study had to
combine five different surveys (one conducted visually and four done
with photographic plates) to increase the number of SN discoveries to
136, but at the expense of introducing heterogeneity. Our study
considered 1036 SNe detected in the KAIT fields, and used an optimal
subset of 726 SNe in the final rate calculation. This is more than
five times the number of SNe used by C99.

(2) {\it Fewer observational biases.} The benchmark work of C99
inherited uncertain biases associated with the limited dynamic range
of human vision and photographic plates.  Our rate calculation, on the
other hand, is based on a single systematic CCD survey with a modern
image-subtraction processing pipeline. Consequently, we are able to
more fully investigate potential observational biases involved in the
rate calculations. In particular, our final rates are based on an
optimal subsample of SNe that excludes the SNe that occurred in
early-type galaxies of small radius (to avoid the uncertainty in the
detection efficiency) and in highly inclined late-type galaxies (to
avoid the uncertain correction factor for galaxy inclination).

(3) {\it Improved quality of the SN sample.} Of the 726 SNe in the
optimal sample, only 12 ($< 2$\%) do not have a spectroscopic
classification. For many of the SNe, we have our own optical
spectra. There are also subclass types for a significant number of
SNe, especially in the complete sample of very nearby SNe (Paper II),
enabling us to provide information on the relative fractions of
different subtypes of SNe.

(4) {\it Fewer uncertainties in the control-time calculation.} The
observed LF in Paper II enables us to avoid several uncertainties that
have historically plagued SN rate determinations, namely the
peak-magnitude distribution and the host-galaxy extinction.  In fact,
our derived LFs should help with all future SN rate calculations.

(5) {\it Improved rate-calculation method.} We identify a trend
between the normalised SN rates and the sizes of the galaxies, and are
able to evaluate the rates in a nominal galaxy size. The rate for a
galaxy of any given size is evaluated with a rate-size relation.

\subsection{Possible Limitations of Our SN Rates} 

There are considerable differences between the LOSS sample galaxy
properties and those from a complete, volume-limited galaxy sample. As
detailed in \S 4, the LOSS galaxy sample has a significant deficit in
faint\footnote{Hereafter, ``faint" refers to low intrinsic luminosity
  unless otherwise specified.} galaxies, especially in the
more-distant bins. This could lead to the belief that our SN rates are
only applicable to luminous galaxies. However, there are three reasons
we think our rates can be applied to the general population of
galaxies, as follows.

(1) Even though the LOSS galaxy sample is not complete at the
low-luminosity end, there is a sufficient number of low-luminosity
galaxies to provide a reliable rate calculation. In fact, the SN rate
in faint galaxies is the cornerstone for the rate-size relation that
we discovered.


(2) From our studies of the SN LFs in Paper II, no significant
differences in the LFs are found for galaxies having different sizes,
with a possible exception for SNe~II in late-type spiral
galaxies. Even in the late-type spiral galaxies, the differences in
the LFs are probably caused by a change in the composition of
different SN subtypes rather than in the luminosities for a particular
SN subtype.

Nevertheless, the possible change of composition for the subtypes of
SNe~II in late-type spiral galaxies, and the growing evidence that
certain kinds of SNe are preferentially found in extremely
low-luminosity, low-metallicity galaxies (e.g., Modjaz et al. 2008;
Miller et al. 2009; Drake et al. 2009; Quimby et al. 2009), lead us to
encourage caution when using our rates in very low-luminosity
galaxies. In the same vein, the reported observed fractions of
different SN subtypes (Paper II) should be used with caution for the
very low-luminosity galaxies. Recently, Arcavi et al.  (2010) reported
a complementary study of a sample of 70 core-collapse SNe found by the
Palomar Transient Factory (PTF), an untargeted survey that monitors
many low-luminosity galaxies; see \S 5.4 of Paper II for more details.

\section{The Control-Time Method for Supernovae with a Known Luminosity 
Function}

The ``control time" is defined as the total interval of time during
which a SN, of a given type and photometric evolution, would have been
bright enough to have been discovered during all of the observations
of a given galaxy.  The use of the control time to calculate the SN
rate was first introduced by Zwicky (1942) and refined by van den
Bergh (1991) and Cappellaro et al. (1993a, 1997).  The idea of being
able to ``control'' a galaxy stems from the fact that SNe are
transient phenomena, and only stay visible for a certain length of
time.

The mathematical details of the control-time method used in our rate
calculations are discussed in the Appendix.  We have generalised the
basic control-time method to the case of a SN type for which we know
the LF. In our study, the LFs are composed of discrete components
corresponding to individual peak magnitudes {\it and light-curve
  shapes} of SNe found in a distance-limited sample, as determined in
Paper II. Our definition of the LF is thus slightly different from the
conventional case where the luminosity function is defined as the
number of sources per unit luminosity interval.

In particular, Equation (A11) is the foundation of our control-time
calculations, and is reproduced here: 

\begin{equation}
t = \sum_{i=1}^{n} f_i t_i,
\end{equation} 
\noindent
where $t$ is the total control time, $t_i$ is the control time from
the $i$-th component in the LF, and $f_i$ is the fraction of the LF
due to the $i$-th component. This equation indicates that the total
control time for a given type of SN with a known LF is the sum of the
control times of each component weighted by the fractional
contribution of that component to the LF.

\section{The Database} 

\subsection{The Galaxy Sample}

\subsubsection{The Construction of the Galaxy Sample}

The LOSS sample galaxies were selected mainly from the Third Reference
Catalog of Bright Galaxies (RC3; de Vaucouleurs et al. 1991). A
detailed description of the selection process is presented elsewhere
(Filippenko et al. 2011).  For the period between March 1998
and the end of 1999, a small sample of $\sim 6000$ fields was used. The
sample was expanded to about 15,000 fields during the year 2000, and
this is the galaxy sample that will be discussed here. We also note
that the sample was enlarged to roughly 20,000 fields during the nearly
three-year period between Oct. 2000 and July 2003 when LOSS was
temporarily expanded to LOTOSS (the Lick Observatory and Tenagra
Observatory SN search; Schwartz et al. 2000).

Although the KAIT camera has a relatively small field of view ($6.7'
\times 6.7'$), there are multiple galaxies in some of the fields. It
is thus important to have a set of criteria to define which galaxies
are monitored during our SN search. As our goal is to conduct detailed
studies of the SN rates in different environments, we require a galaxy
to have the following information to be included in the galaxy sample:
the redshift or distance, the Hubble type, and the luminosity in at
least one of the $B$ or $K$ bands.  We also limit our sample galaxies
to $z < 0.05$ (recession velocities smaller than 15,000 km s$^{-1}$).

To collect information on the KAIT sample galaxies, we have used
exclusively two large online astronomical databases: the NASA/IPAC
Extragalactic Database (NED)\footnote{http://nedwww.ipac.caltech.edu}
and HyperLeda (Paturel et
al. 2003)\footnote{http://leda.univ-lyon1.fr}. We first searched all
of the galaxies in the KAIT fields in NED, and found 76,355
objects. The coordinates for these galaxies were then sent to
HyperLeda to collect additional information. The requirement of having
a recession velocity smaller than 15,000 km s$^{-1}$ cut the sample to
21,190 galaxies; most of the excluded galaxies do not have redshift
information, are visually faint, and are in the background of the
primary KAIT galaxies. We used the recession velocities corrected for
the infall of the Local Group toward the Virgo Cluster (``vvir'' in
HyperLeda), and we adopted $H_0 = 73$ km s$^{-1}$ Mpc$^{-1}$ to
calculate the distance. For a few very nearby galaxies with negative
recession velocities, we were able to find distance estimates using
the VizieR service (Ochsenbein et
al. 2000)\footnote{http://vizier.u-strasbg.fr/viz-bin/VizieR}.

We group the Hubble types of the galaxies into eight bins (Table 1).
These bins can also be grossly labeled as early-type galaxies (No. 1
to 2), early-type spiral galaxies (No. 3 to 5), and late-type spiral
or irregular galaxies (No.  6 to 8). Although on average the
classifications from NED and HyperLeda are consistent with each other
(the difference, using the numbering scheme in Table 1, is $0.15 \pm
0.69$ for 11,754 galaxies that have classifications in both
databases), there is a tendency for HyperLeda to put galaxies having
an Scd classification in NED into the Sc category. Consequently, the
statistics in the Scd galaxy bin become rather poor. We find that the
NED classifications are more consistent with those published in RC3,
from which most of our sample galaxies are selected; thus, we have
adopted the Hubble types from the NED database.

Note that the Hubble-type information in NED comes from many different
sources and is therefore quite inhomogeneous. In fact, there are $\sim
2600$ Hubble-type designations in the database, and we had to
construct a conversion table to put them into our 8 well-defined bins.
We also note that Sydney van den Bergh, together with our group, has
classified the LOSS SN host galaxies in the DDO morphological type
system (van den Bergh, Li, \& Filippenko 2002, 2003, 2005). However,
to perform SN rate calculations using this very homogeneous
classification scheme, we would also need DDO classifications for all
of the KAIT sample galaxies not having a SN discovery, and we were
unable to do this.

For the photometry of the galaxies, we used ``btc" from HyperLeda for
the $B$ band and ``k\_m\_ext" from the 2MASS catalog for the $K$
band. Here, ``btc" is the apparent total $B$ magnitude corrected for
Galactic extinction, internal extinction due to the inclination of the
host galaxy, and the K-correction. The Galactic extinction is adopted
from Schlegel, Finkbeiner, \& Davis (1998). The internal extinction is
computed following Bottinelli et al. (1995), and is a function of the
Hubble type and the axial ratio. The K-correction is calculated from a
simple recipe given by de Vaucouleurs et al.  (1976), and is a
function of the host-galaxy Hubble type and recession velocity. The
value ``k\_m\_ext" is the total $K$-band photometry obtained by
extrapolating the fit of a Sersic function to the radial profile, and
is best suited for extended objects such as galaxies (see Jarrett et
al. 2003, for details).

To calculate the inclinations of the galaxies, which are important
for the investigation of the inclination correction factor for the SN
rate, we follow the traditional Hubble (1926) formula 

\begin{equation}
{\rm Incl} = 
\left\{
\begin{array}{ll}
90^{\circ} & {\rm if~ratio} \le 0.20,  \\
{\rm acos}\Bigl( \sqrt {\frac{{\rm ratio}^2 - 0.20^2} {1.0 - 0.20^2} } 
\Bigr) & {\rm otherwise},
\end{array}
\right.
\end{equation}
\noindent
where ``ratio'' is the ratio of the apparent minor and major axes of
the galaxy.  This simple formula works well for most of the spiral
galaxies, but perhaps not for the irregular and early-type galaxies (van
den Bergh 1988).

The requirement to have the Hubble type and photometry results in
further cuts to our galaxy sample, and our final sample has 14,882
galaxies (hereafter, the ``full" galaxy sample). All of the galaxies in
this sample have Hubble-type information, 98.6\% have $B$-band
photometry, and 91.8\% have $K$-band photometry. The first 20 entries
of this galaxy sample are shown in Table 2, and the rest of the
entries are available electronically. The meaning of the different
columns in Table 2 is as follows: (1) the internal name for the field,
(2) the name of the galaxy, (3) the right ascension ($\alpha$) in
degrees, (4) the declination ($\delta$) in degrees, (5) the distance
in Mpc, (6) the source of the distance, (7) the Hubble type in
numerical code as listed in Table 1, (8) the source of the Hubble
type, (9) the diameter of the major axis in arcmin (isophotal level of
25 mag arcsec$^{-2}$ in the $B$ band), (10) the diameter of the minor
axis in arcmin, (11) the inclination in degrees, (12) the source of
Cols. 9--11, (13) the position angle, (14) the Galactic reddening,
(15) the internal extinction due to galaxy inclination, (16) the total
apparent $B$ magnitude, (17) the uncertainty in Col. 16, (18) the
corrected total apparent $B$ magnitude ``btc," (19) the $K$ magnitude
from 2MASS, and (20) the uncertainty in Col. 19.

Four subsamples of galaxies are used throughout our study.  The
``full" sample has all 14,882 galaxies as discussed above. The second
subsample excludes all of the small early-type galaxies (with major
axis $<1.0'$; see details in \S 4.5), and is called the ``nosmall"
sample. The third subsample excludes all of the small early-type
galaxies and the highly inclined spiral galaxies (${\rm incl.} >
75^\circ$), and is the galaxy sample for the optimal subset of SNe
used for the rate calculations (hereafter, the ``optimal" galaxy
sample). The fourth subsample includes all of the galaxies within
60~Mpc in the ``nosmall" galaxy sample, and is the galaxy sample for
the core-collapse (hereafter, CC) SNe considered in the LF study in
Paper II (hereafter, the ``LF-CC" galaxy sample).  The information on
all of these subsamples is available electronically.

A key question about the LOSS sample galaxies is whether they are
representative of the general population. In the following sections,
we compare the LOSS galaxy LF to the published results from a complete
galaxy sample, and provide statistics on the physical parameters of
the LOSS sample galaxies and their evolution with distance.

\subsubsection{The LOSS Galaxy Luminosity Function}

The upper panel of Figure 1 plots the LOSS galaxy LF for the ``full''
sample. To calculate the absolute $K$-band magnitudes, the $K$-band
magnitudes in Table 2 are corrected for Galactic extinction following
the Galactic reddening law with $R_V = 3.1$ by Cardelli, Clayton, \&
Mathis (1989). No internal extinction due to galaxy inclination is
considered, but it is likely to be minimal. The solid curve is for all
of the galaxies with a $K$-band measurement (13,635 galaxies), while
the dashed line is for 5008 galaxies within 60~Mpc (i.e., the LF-CC
galaxy sample). The LFs are shifted arbitrarily to facilitate
comparison. Also plotted (the smooth solid curve) is the $K$-band
luminosity function for a complete galaxy sample by Kochanek et
al. (2001; the Kochanek curve, hereafter) derived from a sample of
4192 galaxies in the 2MASS survey\footnote{This sample of galaxies
is magnitude limited, but is corrected to be volume limited (complete)
based on the survey volume of each limiting magnitude; see Kochanek
et al. (2001) for details.}, shifted to visually fit the bright
end of the LF-CC sample. The dash-dotted line marks $L_K = 7 \times
10^{10}~ {\rm L}_\odot$ (or $M(K) = -23.7$ mag), which is the nominal
galaxy size used in Paper III.

Compared to the complete sample, there is a clear deficit of
low-luminosity galaxies in the LOSS samples. The LF for the ``full''
sample is dramatically different from the Kochanek curve: the rise at
the luminous end is not as steep, and there is a clear downturn for
galaxies fainter than $M(K) = -24.5$ mag.  However, there is also a
considerable number of low-luminosity galaxies in the LOSS sample:
50\% of the galaxies in the ``full'' sample are fainter than $M(K) =
-23.7$ mag (the nominal galaxy luminosity, about the size of the Milky
Way). 1236 galaxies (8\% of the total) are fainter than $L_K = 0.7
\times 10^{10}~ {\rm L}_\odot$, 10\% of the size of the nominal galaxy
size adopted in Paper III.  Within these low-luminosity galaxies,
26~SNe were discovered and used in the rate calculations.  As shown in
Paper III, despite the incompleteness of the LOSS galaxies at the
low-luminosity end, there are sufficiently high statistics in both the
number of SNe discovered and the control time of the galaxies to allow
a reliable rate calculation to be performed.

For the LF-CC sample of nearby galaxies (distance $< 60$~Mpc),
although the deficit at the low-luminosity end is still clearly
present, the bright end ($L_K > 7 \times 10^{10}~ {\rm L}_\odot$) can
be well fit by the Kochanek curve. This likely means that for these
nearby galaxies, the LOSS galaxy sample is representative of the
complete galaxy sample at the bright end.  This is important,
because even though the LF-CC sample misses a significant fraction of
galaxies at the low-luminosity end (compared to the Kochanek curve) in
terms of numbers, the missing fraction is not significant in terms of
luminosity (and thus in the number of SNe).  This is demonstrated in
the lower panel of Figure 1.  Here the luminosity is plotted on the
abscissa using a linear scale, while the ordinate shows the number of
the galaxies multiplied by the luminosity and the CC~SN
rate\footnote{The rate used here is derived from SNe in all of the
  Hubble types. Also, the rate-size relation in the $K$ band as
  derived in Paper III is employed.}  in Paper III.  Effectively, this
plot illustrates the contribution to the total number of CC~SNe from
each luminosity bin. The contribution peaks at the nominal galaxy
size, has a long tail at the bright end of the galaxies, and a sharp
decline at the faint end. The curve for the complete sample (the
solid smooth curve) is also plotted, and is visually scaled to fit the
bright end of the LF-CC sample. Even though the contribution to the
total CC~SNe peaks at a smaller luminosity ($L_K \approx 4 \times
10^{10}~ {\rm L}_\odot$) for the complete galaxy sample, the
difference in the total number of SNe (the integrals of the two
curves) suggests that the LF-CC sample misses about 15\% of the CC~SNe
due to the deficit in the low-luminosity galaxies relative to the
complete galaxy sample.\footnote{We note that this does not mean that
  our CC~SN rate is off by 15\%, since the rate will be recovered when
  the control time is considered.} A similar analysis is performed
for the SNe~Ia, and only $\sim 10$\% are missed.

We note that these small fractions of missed SNe (compared to the
complete galaxy sample) are consistent with the expectations from a
study by Brinchmann et al. (2004), who suggested that the majority of
the star formation in the local Universe takes place in moderately
massive galaxies, typically in high surface brightness disk
galaxies. The low-luminosity galaxies, though numerous, do not add
much to the total star formation (or total luminosity) and hence do
not produce many SNe. These results have important implications in
Paper II, where we use the SNe discovered in the very nearby galaxies
to construct a complete, volume-limited SN sample.

Although the LF-CC sample is representative of the complete 
galaxy sample at the luminous end, the sample is not complete to all
galaxies within 60~Mpc. First, as detailed by Filippenko et
al. (2011), our survey is limited to declinations between $-25^\circ$
and +70$^\circ$ due to the physical limitation of KAIT, which means we
can only access $\sim 70$\% of the all-sky volume. Since we obtained
snapshot images of a sample of galaxies, the extensive regions in
between the galaxies are not monitored either.  We can estimate the
degree of completeness for our galaxy sample by comparing the total
luminosity to the expectation from the published local luminosity
density.  The total $L_K$ in the LF-CC sample is $1.5 \times 10^{14}~
{\rm L}_\odot$.  Using the local luminosity density as adopted in
Paper III (from Kochanek et al.  2001), the all-sky total luminosity
for the galaxies within 60~Mpc is $L_K = 4.7 \times 10^{14}~ {\rm
  L}_\odot$. This means that our sample galaxies account for $\sim
45$\% of the total stellar light accessible by our survey.

\subsubsection{The Properties of the LOSS Sample Galaxies}

Figure 2 shows some properties of the LOSS sample galaxies. The
upper-left panel illustrates the distribution of Hubble types. The
open histogram is for the ``full'' galaxy sample, while the shaded region
is for the nearby LF-CC sample.  Unfortunately, we could not find a
study of the expected Hubble-type distribution for a complete galaxy
sample, but it seems that the ``full'' galaxy sample has plenty of
galaxies in each Hubble-type bin except the irregular galaxies. This
deficit is most likely caused by the incomplete information for these
visually faint, extended objects in the current astronomical
databases.  In fact, because of the relative lack of ``Irr" galaxies
in the sample, there are just a few SNe discovered in this galaxy bin,
resulting in only an upper limit for many rate calculations.  This is
currently being remedied in our search, as we recently started to
monitor additional LOSS fields centered on known irregular
galaxies. The LF-CC sample (the shaded histogram) has an increasingly
larger fraction of the ``full'' sample from E to Irr galaxies. As
discussed in \S 4.1.4, this is not surprising, as more galaxies of
lower surface brightness are missing from the LOSS sample in
more-distant bins.

The upper-right panel shows the distribution of the inclination angle
for the ``full'' sample. In theory, there should be random lines of sight
for the galaxies, so the distribution for the inclinations in degrees
should follow a sine curve, but there is in fact a dearth of galaxies
having very small inclinations (i.e., face-on galaxies). The reason
for this is the limitation of the catalogued precision for the major
and minor axes (typically $0.1'$). Most of the LOSS galaxies
have a major axis (MaA, hereafter) smaller than $2'$, and require
a precision better than $0.1'$ to yield an inclination smaller
than $20^\circ$. On the other hand, poor precision can also make a
slightly inclined galaxy appear face-on, hence the weak peak at
$0^\circ$ in Figure 2.

The uncertainties in the inclination measurements have some impact on
the results in our study. First, an error in the inclination would
cause an error in the internal extinction calculation to the total
$B$-band magnitude, and this in turn would affect the $B - K$ colour
and the mass estimate for the galaxy.  Fortunately, for galaxies with
inclinations below 20$^\circ$, the internal extinction is small ($<
0.03$ mag). Second, the inclinations are also used to study the
inclination correction factors in the SN rates. For these reasons, we
have grouped the inclinations into three broad bins: the ``face-on"
sample includes all galaxies with inclination smaller than 40$^\circ$,
the ``edge-on" sample is for inclinations larger than 75$^\circ$, and
the ``normal inclination" sample includes all of the rest. The broad
bins are necessary because we want to include a reasonable number of
galaxies in the ``face-on" bin.  Note that edge-on spiral galaxies are
not included in the ``optimal'' galaxy sample.

The lower two panels in Figure 2 show the distribution of MaA in both
arcmin and kpc for the ``full'' galaxy sample. The majority of the
galaxies have MaA between $1'$ and $2'$. The median physical size of
the galaxies is $\sim 25$ kpc.

One main result of our rate calculations in Paper III is that the
normalised SN rate has a correlation with the size of the galaxies. We
thus plot the average $B$ and $K$ luminosity and the mass of the
galaxies in the ``optimal'' sample in Figure 3, and report their values
(together with those for the ``full'' sample) in Table 3. The masses of
the galaxies are calculated according to the prescription of M05 as a
function of $L_K$ and the $B - K$ colour: log$(M/L_K) = 0.212 (B - K)
- 0.959$. Preliminary analysis (Dilday et al. 2011)  of
galaxy spectral energy distribution (SED) fits using PEGASE (Le Borgne
et al. 2004) to multi-band SDSS photometry of a subset of our sample
galaxies indicates that the masses determined using the method of M05
qualitatively agree well with those from the more detailed SED
modeling, suggesting that the M05 prescription is reasonable.  The top
panel of Figure 3 shows the distribution in different Hubble types,
while the bottom panel gives the distribution in different $B - K$
colours. It is clear that galaxies with different Hubble types or
colours have different average sizes (a factor of 10 for the $K$-band
luminosity and mass, and a factor of 2--5 for the $B$-band
luminosity).  Not surprisingly, the most massive galaxies are the red
ellipticals.

\subsubsection{The Evolution of LOSS Sample Galaxies with Distance}

Although no cosmological evolution effect is expected for the nearby
LOSS galaxies, we demonstrate in this section that there are important
systematic changes in the galaxy properties due to selection biases
over the 0 to $\sim 200$~Mpc distance range of our sample.

Figure 4 illustrates the distribution of several parameters as a
function of distance. The upper-left panel shows the $K$-band
luminosity of the ``full'' galaxy sample; each dot represents a
galaxy. Also plotted is the average $K$ luminosity in different
distance bins. The dots show that at greater distances, low-luminosity
galaxies tend to be missing.  Consequently, the average $K$ luminosity
increases monotonically with increasing distance. This is due to a
Malmquist bias; the current astronomical databases become increasingly
incomplete for low-luminosity objects at greater distances.  As
discussed in Paper III (especially in its Appendix), the presence of
this obvious Malmquist bias is the main reason for our discovery of a
strong correlation between the SN rates and the sizes (luminosities or
masses) of the host galaxies (i.e., the rate-size relation).

The lower-left panel shows the average $K$ luminosity of galaxies in
different distance bins, for two Hubble types (open circles for Scd,
solid circles for E). The distribution of the E galaxies has been
scaled to match the Scd distribution in the 0--50 Mpc range. It is
evident that the Scd galaxies exhibit a more dramatic Malmquist bias
than the E galaxies, due to their low surface brightness and the
general absence of bright nuclei. The average $K$ luminosity changes
by a factor of 20 over the distance range 15 to 175 Mpc for the Scd
galaxies, while the corresponding factor for the E galaxies is 4.

The upper-right panel displays the number of galaxies in different
distance bins. From 0 to 60~Mpc, the number of galaxies per bin
increases, as one would expect from a complete galaxy sample; the
survey volume per bin progressively increases with distance. However,
the distribution peaks at 60--70 Mpc, and then declines gradually at
greater distances.

The lower-right panel shows significant differences in the number
distribution with distance of two types of galaxies (solid line for E,
dashed line for Scd).  While the number of E galaxies per bin
increases from 0 to 60~Mpc, stays nearly constant until 140~Mpc, and
declines thereafter, the number of Scd galaxies per bin is almost
constant from 0 to 80~Mpc and then declines sharply thereafter. Again,
this reflects completeness differences in the current astronomical
database for galaxies of different Hubble types. The low surface
brightness Scd galaxies are only complete nearby, while the E galaxies
are easier to observe and are complete to a greater distance. 

Because galaxies of different Hubble types have different
distributions with distance, the relative fractions of the Hubble
types also change: there are more late-type spirals in the nearby
distance bins, while the E galaxies dominate the most distant bins.
The change in Hubble-type demography with distance has important
implications for the SN rates and will be discussed throughout our
study.

\subsection{The SN sample}

\subsubsection{The Construction of the SN Sample}

LOSS began in 1997 but found only a single SN of questionable nature
during that year (SN 1997bs; Van Dyk et al. 2000). Numerous
improvements were made during late 1997 and early 1998, including the
replacement (in March 1998) of the original Photometrics CCD camera
having a front-illuminated Thompson TH 7895 chip with an Apogee AP7
having a much more sensitive back-illuminated SITe 512 chip.  For our
SN rate calculations hereafter, we consider only the period March 1998
through December 2008 (about 10 years and 9 months), unless explicitly
expressed otherwise.

To determine accurate SN rates, we need to consider not only the SNe
found by LOSS itself, but also the ones first discovered by other
groups which were subsequently found independently as part of our
search. Although we keep log files of the SNe discovered in the LOSS
fields, to ensure completeness, we cross-correlated the official list
of all the discovered SNe on the CBAT
website\footnote{http://www.cfa.harvard.edu/iau/lists/Supernovae.html}
with the LOSS fields, and generated a list of 1232 SNe that were
within the field of view of these fields. After setting aside the 732
SNe first discovered by LOSS, we then went through the monitoring
history of the fields of the other 500 SNe, checked the
image-processing log files and the candidate reports, examined the
reobservation history, and identified 304 SNe that were found
independently during our search. Thus, LOSS discovered a total of 1036
SNe.

Of the 196 SNe that were missed by LOSS, (a) the vast majority were
discovered by other searches at a time when KAIT was not actively
monitoring their host galaxies (fields that are too far toward the
west at the beginning of the night, or too far toward the east at the
end of the night); (b) some are background SNe at higher redshifts in
the fields of the targeted LOSS galaxies; and (c) three were very
close to the nuclei of their host galaxies (SN 2002bs, Wei et
al. 2002; SN 2004cm, Connolly 2004; SN 2006gy, Quimby 2006) and
not detected by the LOSS image-processing software (see additional
discussion in \S 4.5). None of the objects detected by the pipeline
was overlooked due to human error (see \S 4.3).

We cross-correlated the 1036 SNe that were found directly or
independently by LOSS with the ``full'' galaxy sample and selected 934 SNe
for use in the rate calculations. Of the remaining 102 SNe that are
within the LOSS fields but not considered in the rate calculations, 47
are in background galaxies (discussed further in Paper II), 19 have no
or incomplete Hubble-type information for their host galaxies, and 36
have no known host-galaxy redshift or $z > 0.05$.

Our final total SN sample consists of 929 SNe, after five additional
SNe were removed as follows. SN 2003dl (Graham \& Li 2003) was
reported as a LOSS discovery, but a careful reanalysis of the
monitoring images using better template images than were available at
the time of discovery suggest that there is no SN at the position of
SN 2003dl; instead, the detection was likely caused by a random
background fluctuation combined with an inferior template image. SN
2005md (Li 2005) was also discovered by LOSS. Modjaz et al. (2005)
obtained a spectrum, and suggested the object to be similar to the
SN~IIb 1993J at early times based on its featureless, blue
continuum. However, the SN nature of SN 2005md is now highly
questionable, given that it rebrightened in 2008 (Li et al. 2008). SN
2005ha (Prasad et al. 2005) was discovered by LOSS in the nearby S0/a
galaxy UGC 3457 ($D = 36$ Mpc). It was quite subluminous at peak
brightness ($-14.8$ mag) and had a fast photometric evolution. A
spectrum obtained by Hamuy, Maza, \& Folatelli (2005) is inconclusive,
showing only a hint of the Ca~II near-infrared triplet in emission. It
is possible that SN 2005ha belongs to the growing subclass of
subluminous ``Ca-rich SNe~Ibc'' (Filippenko et al. 2003; Perets et
al. 2010). We will provide more analysis of SNe 2005md and 2005ha in a
future paper, but for now they are not considered in the rate
calculation.  SNe 2002bj and 2004cs are also transients of uncertain
identity, as discussed below, and were removed from the sample.

One of the great assets of our SN sample is the completeness of the
spectroscopic classification: useful spectra were obtained of 917 out
of the 929 SNe (98.7\%), thanks to the dedicated efforts of several
groups. In particular, the majority of the classifications were made
with the Mt. Hopkins 1.5~m telescope by the SN group led by R.~P.
Kirshner at the Harvard-Smithsonian Center for Astrophysics.  Many of
the classifications were also made with the Lick Observatory 3~m Shane
reflector by our own SN group (led by A.V.F.) at University of
California, Berkeley. We compiled the spectroscopic classifications
from several sources and checked for differences: (a) the official
list of SNe on the CBAT website as mentioned above, which contains the
classifications; (b) the Asiago SN
catalog\footnote{http://web.oapd.inaf.it/supern/cat/}; (c) the SN list
maintained by
M.~W. Richmond\footnote{http://stupendous.rit.edu/richmond/sn.list};
and (d) a list of SNe and their classifications compiled by us over
the years.  When there were discrepant classifications, we consulted
the original IAU Circulars and our spectroscopic database. In the end,
we have self-consistent classifications for all of the
spectroscopically observed SNe in the sample; we believe there is only
a low level ($\sim 2$\%) of misclassifications.

In Paper II, we fit the light curves of the SNe in the LF sample, thus
providing a consistency check for the spectral classifications. This
process led to the following four revisions or removals.

\begin{enumerate}

\item{Filippenko \& Chornock (2002) classified SN~2002au as a probable
  (but not definite) SN~Ia. The light-curve fit in Paper II suggests
  that the object is instead a SN~IIb. We analysed the spectrum
  observed by Filippenko \& Chornock using the Supernova
  Identification code (SNID; Blondin \& Tonry 2007), and indeed the
  best three spectral matches are all SNe~IIb near maximum brightness. }

\item{Serduke et al. (2006) classified SN 2006P as a probable SN~Ia at
  about 2 weeks past maximum brightness. The analysis of the light
  curve in Paper II instead suggests a SN~Ibc classification.  We
  again used SNID to analyse the spectrum, and the best three matches
  are all SNe~Ic near maximum brightness.}

\item{SN 2002bj was classified as a SN~II or IIn by Kinugasa et
  al. (2002).  Reduction of our follow-up photometry reveals a
  peculiar light curve: the SN declined by $\sim 5$ mag over a short
  period of 15~d, faster than any other known SN (Poznanski et
  al. 2010). The two spectra we obtained of SN 2002bj do not exhibit
  obvious H Balmer lines, and show some resemblance to the peculiar
  SN~Ibc~2006jc (Foley et al. 2007; Pastorello et al. 2007). On the
  other hand, Poznanski et al. (2010) argue that SN 2002bj may have
  been a ``.Ia'' supernova (Bildsten et al. 2007), or at least some
  kind of strange partial explosion of a white dwarf. Given its
  uncertain classification, we have removed SN 2002bj from our present
  analysis and will consider it further in a future paper.}

\item{SN 2004cs (Li, Singer, \& Boles 2004) does not have a
  spectroscopic classification published in the IAU Circulars, but
  Rajala, Fox, \& Gal-Yam (2004) suggested that the object was a SN~Ia
  based on its observed colours and the colour-typing method developed
  by Poznanski et al. (2002), Gal-Yam et al. (2004), and Poznanski,
  Maoz, \& Gal-Yam (2007).  We have an excellent unfiltered light
  curve of SN 2004cs which reveals rapid evolution not compatible with
  that of any observed SNe~Ia. Analysis of a spectrum (A. Gal-Yam
  2009, private communication; also illustrated by Rajala et al. 2005)
  indicates that SN 2004cs is similar to SN 2007J, which may have been
  a peculiar SN~Ibc (Filippenko et al. 2007) or perhaps related to a
  certain odd subclass of SNe~Ia (Foley et al. 2009).  As with SN
  2002bj, we have removed SN 2004cs from our present analysis and will
  consider it further in a future paper.}

\end{enumerate}

SNe are spectroscopically classified into three main categories: Ia,
Ibc\footnote{Here we use ``Ibc'' to generically denote the Ib, Ic, and
  hybrid Ib/c objects whose specific Ib or Ic classification is
  uncertain.}, and II, with subclasses in each category (for a review,
see Filippenko 1997).  While we concentrate our rate calculations on
the three main classes, we also pay attention to the subclasses,
especially for the SNe in the LF sample (see Paper II for
details). SNe~Ia are grouped into the following subclasses: (a) normal
objects (e.g., Branch \& Tammann 1992), (b) high expansion velocity
objects (Benetti et al. 2005; Wang et al. 2009), (c) SN 1991T-like
objects (Filippenko et al. 1992a; Phillips et al. 1992), (d) SN
1991bg-like objects (Filippenko et al. 1992b; Leibundgut et al. 1993),
and (e) SN 2002cx-like objects (Li et al. 2003; Jha et al. 2006). The
SN~Ia nature of SN 2002cx-like objects is currently being debated
(Valenti et al. 2009 --- but see Foley et al. 2009, 2010), and their
rate will be further discussed in a future paper.  SNe~Ibc are
classified into (a) SNe~Ib, (b) SNe~Ic, (c) peculiar SNe~Ibc
(hereafter, Ibc-pec), and (d) SNe~Ib/c (SNe~Ib or Ic with uncertain
subtype classification). SNe~II are classified into (a) SNe~II-P
(``plateau'' in the light curve), (b) SNe~II-L (``linear'' magnitude
brightness decline after the peak), (c) SNe~IIb, and (d) SNe~IIn. Note
that the classifications of SNe~II require both spectroscopic and
photometric information, and are thus possible only for a subset of
the SNe in our sample (the LF SNe in Paper II).

Information for the SN sample is listed in Table 4. Only the first 20
entries are shown, with the rest available electronically. Each SN
entry lists the following: the SN name, its host-galaxy name, the date
of discovery, the right ascension ($\alpha$) in degrees, the
declination ($\delta$) in degrees, the offsets (in arcsec) from the
host-galaxy nucleus, the magnitude at the time of discovery, the
classification, the discoverer, and the membership in the various
subsamples in our SN rate analysis (discussed below in \S 4.2.2). 
Among the 929 SNe in the total SN sample, 372 are SNe~Ia
(40.0\%), 144 are SNe~Ibc (15.5\%), 399 are SNe~II (42.9\%), and 14
have no spectroscopic classification (1.5\%).  For the 726 SNe in the
``optimal'' sample used in the final SN rate calculations, 274 are SNe~Ia
(37.7\%), 116 are SNe~Ibc (16.0\%), 324 are SNe~II (44.6\%), and 12
have no spectroscopic classification (1.6\%).  For the rate
calculations, SNe without a spectroscopic classification are split
into fractions of SNe~Ia, Ibc, and II according to the statistics of
the SNe having spectroscopic classifications.  These fractions are
also different from the observed fractions in a complete sample of
very nearby SNe as discussed in Paper II.

In the following sections, we describe several aspects of our SN 
sample in an attempt to quantify the characteristics of the sample. 

\subsubsection{The Different SN Subsamples}

Our study uses a relatively large number of subsamples, seven total,
in order to thoroughly explore various issues.\footnote{The list of
  SNe in each of the various subsamples is available electronically.}
The ``full" sample includes all 929 SNe discussed above. The
``full-nosmall" subsample excludes the SNe in small early-type
galaxies (MaA $< 1'$), and has 884 SNe. The ``full-optimal" subsample
excludes the SNe in small early-type and edge-on spiral galaxies, and
has 726 SNe.

In an attempt to alleviate uncertainties in the control-time
calculations due to uncertainties in the LFs and the host-galaxy
extinction distribution, we consider a subsample of SNe that were only
discovered ``in season" --- that is, the SNe exploded during (not
prior to) the active monitoring period of the galaxies.  We set a
simple criterion for a SN to be considered ``in season": there had to
be a nondetection deeper than the SN discovery magnitude shortly
before the discovery was made. In other words, a SN discovered in the
first image of a galaxy after a long break when the galaxy was too
close to the Sun was not counted as an ``in-season SN.''  Accordingly,
for this subsample the control time for each galaxy does not consider
the first image after a long break. Note that because of our small
observation intervals, this first image was often the only instance
when the control time even needed to be calculated using the
light-curve shape and limiting magnitude (\S 4.3); thus, we largely
eliminated the control time from the calculation.

We went through the monitoring history of all 929 SNe and identified
656 ``in-season" SNe, which we call the ``season" subsample.  The
``season-nosmall" subsample excludes the SNe in small early-type
galaxies and has 617 SNe. The ``season-optimal" subsample excludes the
SNe in small early-type and edge-on spiral galaxies; it contains 499
SNe.

After we constructed the observed LFs (Paper II) and used them in our
SN rate calculations, the necessity of using the various season
subsamples is diminished. Nevertheless, these subsamples offer us a
chance to compare the rate measurements from different subsamples, as
we have done in Paper III. Compared to the ``full'' samples, the ``season''
subsamples sacrifice some SNe and thus are more susceptible to
small-number statistics, but provide better tolerance to the
uncertainties in the LFs.

A subset of 175 SNe in the ``season-nosmall'' subsample (CC~SNe within
60~Mpc, and SNe~Ia within 80~Mpc), which we call the ``LF" subsample,
is used to construct the LFs in Paper II.

\subsubsection{The Hubble-Type Distribution of the SN Host Galaxies}

Figure 5 shows the Hubble-type distribution for the host galaxies of
the ``full" SN sample. There is a significant difference between the
distribution of the SN~Ia hosts and that of the CC~SN hosts, as we
have previously reported (van den Bergh et al. 2002, 2003, 2005).
Kolmogorov-Smirnov (hereafter, K-S) tests suggest that there is only a
$2.0 \times 10^{-9}$ and $1.0 \times 10^{-20}$ probability that the
SN~Ia hosts come from the same population as the SN~Ibc and SN~II
hosts, respectively. Even when the early-type galaxies (E--S0) are not
considered, the SN~Ia hosts still have a significantly different
distribution compared with that of the CC~SNe; the probability that
they come from the same population is less than 2\%. On the other
hand, the hosts of SNe~Ibc and SNe~II have quite similar
distributions, coming from the same population with 94.9\%
probability.

These results are not surprising, given the different progenitor
systems for the different SN types. SNe~Ia are generally thought to
come from the explosion of an accreting white dwarf in a binary
system, so they are frequently found among old or intermediate-age
populations of stars. Core-collapse SNe (Ibc and II), on the other
hand, come from stars that are more massive than 8--10~M$_\odot$, so
they arise predominantly from young, star-forming populations. When
translating this correlation into the Hubble-type distributions,
SNe~Ia are often found in E, S0, and early-type spiral galaxies, while
SNe~Ibc and II occur mostly in spiral galaxies.

There is a small fraction (13 out of 536; 2.4\%) of CC~SNe in the
early-type galaxies in our SN sample, as listed in Table 5. Hakobyan
et al. (2008) carried out a detailed morphological study and an
extensive literature search for a sample of 22 CC~SNe that had
apparently occurred in early-type galaxies, and found a significant
fraction (17 out of 22) of these galaxies to actually be misclassified
spirals. They also discovered that for the genuine early-type
galaxies, there are independent indicators of the presence of recent
star formation due to mergers or interactions. As listed in Table 5,
the majority of the early-type galaxies with CC~SNe in our sample are
``S0/a," which is in between S0 and Sa and should have a small amount
of recent star formation. The host galaxy of SN 2007ke is classified
as ``E" in both NED and HyperLeda, but it is interacting with another
elliptical galaxy and is a member of a bright cluster of galaxies. The
host galaxy of SN 2003ei is classified as ``E" in NED, but it occurred
in a tidal arm of an interacting galaxy pair.  The host galaxy of SN
2005ar is classified as ``E" in both NED and HyperLeda; however, we
inspected several images from KAIT and DeepSky (Nugent 2009), and
suggest that this may be another misclassification because the galaxy
does not have a bright nucleus as seen in classical elliptical
galaxies and appears to have some diffuse spiral-arm emission.

Thus, we conclude from this exercise that CC~SNe in ``E" and ``S0"
galaxies (our galaxy bins 1 and 2) are rare, especially after
excluding the ones in ``S0/a" galaxies and the misclassifications. The
scarcity of CC~SNe in the early-type galaxies places a strong limit on
the amount of recent star formation within these galaxies, which will
be used in Paper III when we discuss possible causes for the observed
rate-size relation.

We also note the predominance of SNe~Ibc over SNe~II in Table 5: of
the total 536 CC~SNe, only 144 are SNe~Ibc (27\%), but 7 out of 13
(54\%) of the CC~SNe in the early-type galaxies are SNe~Ibc.  In
particular, there is a strong preference for the so-called ``Ca-rich
SNe~Ibc'' (Filippenko et al. 2003; Perets et al. 2010) to occur in
early-type galaxies.  About 10 Ca-rich SNe have been identified, all
of which were discovered in the LOSS galaxies, and three of them have
relatively early-type hosts.  As discussed by Perets et al. (2010),
the association of Ca-rich SNe~Ibc with early-type galaxies provides
clues to the nature of their progenitors: they probably represent a
new type of stellar explosion arising from a low-mass and relatively
old stellar system.

\subsubsection{The SN Distribution as a Function of Distance}

The left panels of Figure 6 show the distribution as a function of
distance for the different types of SNe. SNe~Ia, because of their
extraordinarily high luminosity, are typically discovered at much
greater distances than the CC~SNe.  The distributions for the SNe~Ibc
and II are rather similar, reaching a peak at around 50--70 Mpc and
displaying a sharp decline after 100--110 Mpc.

The distribution over distance depends on several factors,
particularly the SN luminosity, the distribution of Hubble type and
galaxy size with distance, and the control time for each
galaxy. Because each distance bin has roughly 1000 galaxies, the
average control time might be similar in the very nearby distance bins
where we have total control during the observing seasons, and then
decline in more distant bins when we have only partial control. It is
thus possible to estimate at which distance our SN survey has full
control for the different SN types.

The right-hand panel of Figure 6 shows how the number of SNe in each
distance bin, divided by the total $K$-band luminosity for {\it all}
galaxies in that bin, evolves with distance.  The curves are visually
shifted to approximately match the first several distance bins. We
expected the curves to show a constant for the nearest distance bins
(where the average control times are the same for the different
distance bins because we have full control), and then drop for the
more distant bins. But instead, the curves exhibit an apparent decline
even for the most nearby distance bins. This is the first sign of the
presence of a size dependence for the SN rates: even though the
average control time may be the same for the very nearby distance
bins, the average galaxy size increases significantly with increasing
distance (Figure 4), so the average SN rate declines with distance.

Nevertheless, there seems to be a clear divergence between the SN~Ia
and CC~SN curves starting at $\sim 70$ Mpc. Since SNe~Ia are more
luminous than the CC~SNe, we should have full control of SNe~Ia over a
larger distance; hence, we take the divergence as a sign that the
control time for CC~SNe begins to fall short of the total observing
season time at distances beyond 70~Mpc. For this reason, and to be
conservative, in Paper II we elect to construct our LF sample of
CC~SNe using a distance cutoff of 60~Mpc (dashed line in Figure 6).

\subsubsection{The Radial Distribution of SNe}

Our search is conducted with a CCD camera and the SNe are discovered
via image subtraction, so the discrimination against SNe occurring
near the bright nuclei of galaxies is not as severe as in surveys with
photographic plates.  Nevertheless, we exclude an area with a radius
of a few pixels (3--4, depending on the seeing, corresponding to
$2.4''$--$3.2''$) centered on every galaxy nucleus during our search,
because galactic nuclei often suffer imperfect image subtraction and
introduce many false SN candidates. In order to estimate the fraction
of SNe missed as a result of this SN search strategy, we determined
the surface density of SNe as a function of radial distance
from the center of the host galaxy. In \S 4.5, we will also perform
Monte Carlo simulations to achieve the same goal.

The SNe in our sample have well-documented offsets, $x\arcsec$ east or
west and $y\arcsec$ north or south of their host-galaxy nuclei. Under
the assumption that the galaxies are circular disks and only appear to
have different major and minor axes due to their inclination, we can
calculate the radial distance of the SNe ($R_{\rm SN}$) from the
nuclei if we know the position angles and the axis ratios of the
galaxies. The radius of the galaxy ($R_{\rm gal}$) is simply half of
the major-axis diameter. The ratio $\theta = R_{\rm SN}/R_{\rm gal}$
is then the fractional radial distance for the SN, and it may be
compared for different objects.  We group $\theta$ in different bins,
and calculate the surface density as $N_i/[\pi (\theta_i^2 -
\theta_{i-1}^2)]$, where $N_i$ is the number of SNe in bin $i$.

Figure 7 displays the surface-density distribution versus the fractional
radial distance $R_{\rm SN}$/$R_{\rm gal}$. The top panel shows the
distribution of two SN groups, one with the MaA of the host galaxy
greater than $1.25'$ (a total of 433) and the other with smaller
host galaxies (a total of 373). The split is designed so that the two
distributions are similar for $\theta > 0.3$. The SN distribution in
bigger galaxies illustrates that the occurrence of SNe follows an
exponential trend, with the highest density in the central regions of
galaxies. This is not surprising; the SN occurrence should follow the
same trend as the stellar population density in galaxies.  The SN
distribution in the smaller galaxies, however, shows an apparent
change starting at $\theta = 0.25$, and becomes consistently lower at
small $\theta$ than the measurements for the bigger galaxies.  This
difference is likely caused by the missing SNe in the central region
of the smaller galaxies. Taking the difference in the number of SNe
(60) in the two groups as a rough estimate of the missed SNe, about
$60/810 \approx 7$\% of the SNe were missed in our sample. Since our
survey likely missed a small fraction of the SNe in the center of the
bigger galaxies, the total fraction of SNe missed by our survey is
probably $\sim 10$\%.  We note that this missed SN fraction is similar
to what we derive from Monte Carlo simulations in \S 4.5 when the
detection efficiency of our survey is estimated.  We also emphasise
that by implementing detection efficiency in the control-time
calculation, the missed SN fraction is taken into account when the
final rate calculations are performed.

We note that the majority of the $\sim 60$ SNe missed in our search
were not discovered by other SN search groups either. As discussed in
\S 4.2.1, only three reported SNe in the LOSS sample galaxies were
missed in our survey because they occurred too close to the nuclei of
their host galaxies. Thus, either we have overestimated the number of
missed SNe in our search, or all current SN searches suffer some degree
of incompleteness for the SNe in bright galactic nuclei.

It is of interest to check whether the different SN types have
different surface-density distributions. The lower panel of Figure 7
shows three curves for SNe~Ia, Ibc, and II that are visually shifted
to match each other. One notable difference is that there is a dip at
small $\theta$ values in the SN~II distribution. Perhaps SNe~II have a
higher missed fraction near galaxy nuclei compared to the other two SN
types, or SNe~II prefer not to occur near galaxy nuclei. On the other
hand, SNe~Ibc seem to be more concentrated within $\theta < 0.70$
(86\% of all) than the other SN types (76\% and 71\% of all for SNe~Ia
and II, respectively). To evaluate the significance of these
differences, we ran K-S tests on the $\theta$ distribution of the SNe,
and show the cumulative fractions in Figure 8. Significant differences
are found between SNe~Ibc and SNe~Ia/II, with a respective probability
of 2.9\% and 1.4\% that the SNe come from the same radial
distribution.

 We checked the above results using SNe in relatively big
galaxies (in this case, MaA $> 1.0'$ to maintain a reasonably large
sample) to alleviate the effect of missed SNe in the centers of
galaxies, and the K-S test probability is (respectively) 9.7\% and
5.6\%, so there is still a significant difference between SNe~Ibc and
II.  The fact that SNe~Ibc prefer to occur at $\theta < 0.70$ possibly
indicates a bias toward formation of their progenitors in the higher
metallicity regions of galaxies, consistent with what is inferred from
the host-galaxy property study in \S 5.4 of Paper II, where SNe~Ibc
are found to preferentially occur in more massive (and thus higher
metallicity) galaxies than SNe~II.  We note that a similar metallicity
dependence between SNe~Ibc and II is reported by Boissier \& Prantzos
(2009); see also Prantzos \& Boissier (2003) and Prieto, Stanek, \&
Beacom (2008).

We also split the SN~Ibc sample into SN~Ib and SN~Ic subsamples, and
show the cumulative fractions of their radial distributions in Figure
8. While no significant difference is found among the SN~Ib, Ic, and
Ibc samples, the SN~Ic sample boasts the most significant difference
in radial distribution from the SN~II sample; the two samples come
from the same population at only 0.1\% probability, compared to 15.4\%
between the SN~Ib and SN~II samples.  The SN~Ic sample seems to have
the highest central concentration among the core-collapse SNe. This is
consistent with the results of Kelly, Kirshner, \& Pahre (2008), where
SNe~Ic were found to have the most significant difference in spatial
distribution compared to other core-collapse SNe.

Since the radial distance is only a rough estimate of the location of
a SN in its host galaxy, more sophisticated methods are necessary to
reveal additional differences in the distribution of the different
type of SNe in their host galaxies (e.g., Fruchter et al. 2006; Kelly
et al. 2008).

\subsection{The Log Files}

Numerous log files are assembled as part of our SN search. The
telescope observing log files record the details of each observation,
including weather conditions, observing parameters such as hour angle
and declination, and possible problems. The image-processing log files
record all of the details regarding image subtraction and candidate
detections.

SN-candidate log files keep track of the history of each SN discovered
by our own search, or SNe that were discovered by other groups first
but were found by our survey independently at a later time. Part of
the goal for keeping these candidate log files is to check whether
some of the SN candidates found by the image-processing software are
missed by students in our team during the human image-checking
process. As detailed in \S 4.2.1, 304 SNe were independently
discovered during our survey, and none of the SNe was missed during
the image-checking process. Occasionally, a student would miss a SN
candidate during the image checking for a single night, but because we
have a short observation interval, the field would be repeated several
times with the SN still visible, and it would eventually be noticed by
the same student or another student. So, overall, our survey has not
missed a single SN due to human error.

Another valuable log file is the observing history of each individual
field.  Each entry in this log file records the date of the
observation, the camera used, the adopted template image, and other
potentially relevant information about the image that can be used to
derive its limiting magnitude (see details in the next section): the
intensity ratio relative to the template image, the sky background,
and the seeing as measured by the full width at half-maximum intensity
(FWHM) of the stellar images.

One important statistic is the average observation interval for our
sample galaxies. A histogram of 2.3 million observations (with typical
exposure time of 16~s to 20~s) considered in this rate calculation is
shown in Figure 9.\footnote{We do not consider the long interval when
  a galaxy is too near the direction to the Sun for observations.}
The average observation interval is 8.7~d, and the majority ($\sim
73$\%) of the intervals are smaller than 10~d. The histogram also
shows a bimodal distribution, with peaks at around 5~d and 9~d. This
is the result of our survey strategy: a few hundred very nearby
galaxies were set to repeat every 2~d, about 5000 galaxies were set
to repeat every 5~d, and the remaining 10,000 galaxies were set to
repeat every 10~d. The small observation interval of our search is
a major reason why our rates are relatively insensitive to the adopted
LF in the control-time calculation, especially for the luminous class
of SNe~Ia, as discussed in Paper III.

\subsection{The Limiting-Magnitude Estimates}

The limiting magnitude of the survey images is an important ingredient
for the calculation of control times. Historically (e.g., C99), the
limiting magnitude was often set as a constant for a particular
survey, but in reality, the limiting magnitude for each image is
different due to different observing conditions such as clouds, the
seeing, and the sky background. We have all of the individual survey
images, so in theory we could measure the limiting magnitude for each
image, but the data-processing time for over 2 million images is
prohibitive. Instead, we derive empirical correlations between the
limiting magnitudes for a subset of the images and several parameters
recorded in the log files, and apply these correlations to the
remaining images.
 
We find a tight correlation between the limiting magnitude and the
logarithm of the intensity ratio (of the search image to the template),
the FWHM of stars, and the logarithm of the sky background. We also
find that the correlations are different for different combinations of
the CCD cameras used to take the image and the template. During the
period of our survey that is considered in this rate calculation, we
have used three CCD cameras: an Apogee AP7 camera with enhanced
ultraviolet coating between March 1998 and Sep. 2001, an Apogee AP7
camera with a broad-band coating between Sep. 2001 and May 2007, and a
Finger Lakes Instrumentation Proline Camera after May 2007. There are
5 combinations of the cameras used to take the new image and the
template.

For each combination, we choose a large number of images
(1000--40,000) that were taken under different weather and observing
conditions, and derive their limiting magnitudes as follows.  All of
these images are selected from fields that have either reliable
calibrations from our follow-up campaign (Ganeshalingam et al. 2010),
or many stars that are calibrated in the USNO B1 catalog (Monet et
al. 2003). For each image, artificial stars, with known brightness
determined from the calibration and point-spread function (PSF)
constructed from bright, isolated stars in the same image, are
randomly injected in the image. We then use the ``sextractor program''
(Bertin \& Arnouts 1996) to determine how many of these artificial
stars are recovered using a set of default parameters that are also
adopted for the candidate detection in the SN search software. We
repeat this process with increasingly fainter magnitudes for the
artificial stars until $>50$\% of them are undetected, at which point
we set that magnitude as the limiting magnitude of the image. Note
that our particular definition of the limiting magnitude is not
important, because it is also adopted in our Monte Carlo simulations
to determine the detection efficiency, which is eventually used in the
rate calculations.

After all of the limiting magnitudes are derived, we use a
multiple-variable-regression program to derive the coefficients for
the three components mentioned above. Overall, we achieve a solution
with a scatter of 0.2--0.3 mag for over 50,000 limiting
magnitudes. Figure 10 shows the correlation between the limiting
magnitude and the intensity ratio, after the corrections for the other
two components (stellar FWHM and sky background) have been made. The
overall limiting magnitudes for our survey images spread over a wide
range, but are concentrated between 18.0 and 19.5, with a median value
of $18.8 \pm 0.5$ mag.

\subsection{The Detection Efficiency}

We define the detection efficiency (DE, hereafter) of an image in our
survey as the probability of detecting a SN-like point source in that
image as a function of the difference between the SN brightness and
the limiting magnitude of the image as derived in \S 4.4.  One could
assume a step function for the DE (1.0 when the SN is brighter than
the limiting magnitude, 0.0 otherwise), but that is only a rough
approximation of the real situation; SNe are more likely to be lost in
the bright nuclear regions of galaxies, and the limiting magnitude
determined in \S 4.4 is most suitable for isolated stellar sources,
not for SNe that are generally superimposed on a host-galaxy
background. For this reason, we perform Monte Carlo simulations to
determine the DE in our survey, similar to what has been done
previously (e.g., Pain et al. 2002; Gal-Yam, Maoz, \& Sharon 2002;
Blanc et al. 2004; Neill et al. 2006; Sullivan et al. 2006; Sharon et
al. 2007).

In an attempt to investigate whether the DE is correlated with the
properties of the galaxies, we select 34 galaxies having different
Hubble types and sizes.  For each galaxy, we choose several images
that were observed under different observing conditions so that we
have good coverage of the limiting magnitudes. A total of 189 images
are selected to participate in the simulations. Additional images of
more galaxies are in principle desirable to sample the total range
of galaxy morphology and observing conditions, but the DE simulation
is a time-consuming process (in both computation and analysis), and we
reach the point of diminishing returns as the DE curves converge
toward the same galaxy Hubble type (see discussion below).

We follow the prescription of Gal-Yam et al. (2002) to inject
simulated SNe into the images, with a spatial distribution that
follows the flux of the galaxy. For this purpose, we construct deep
template images for each of the 34 galaxies by stacking the images
observed under the best conditions. Stars that are within the galaxy
profiles are carefully removed by modelling their PSFs. The remaining
flux maps of the galaxies are used as the probability maps where the
simulated SNe should be located. We emphasize that the simulated SN
injection process does not avoid the nuclear region of a galaxy.
Quite the contrary, the nuclear region usually has the highest flux of
the galaxy so it has the highest possibility (per unit area) of
harboring a simulated SN.

For each image, we study the detection efficiency for SNe that are
from 2~mag brighter to 1~mag fainter than the limiting magnitude,
using a 0.2~mag increment. For each SN magnitude, 20 simulated SNe are
randomly injected into the image one at a time according to the flux
map.  The image is then processed with the same software used in our
SN search to do template image subtraction and SN candidate detection.
As in the case of our actual search, the central few pixels in the
nuclear region of a galaxy are excluded.  The SN candidate positions
are checked against the input coordinates of the simulated SNe to
determine whether they are recovered.  The efficiency derived in this
manner then naturally accounts for parts of the image that are not
useable for the SN search, such as the central several pixels in the
bright galaxy nucleus (because the simulated SNe are injected there
according to the galaxy flux maps, but the image-processing software
excludes these regions in both our actual search and the simulations).

Our DE results are shown in Figure 11. We find that the DE curves do
not change significantly for different limiting magnitudes, but show a
strong dependence on the Hubble types of the galaxies. The curves are
flat when the SN magnitude is brighter than the limiting magnitude by
more than 1~mag, then have a dramatic drop when the SN is 0.5~mag
fainter than the limiting magnitude, and reach zero when the SN is 
1~mag fainter than the limiting magnitude. At any given SN magnitude,
the early-type galaxies (E--S0) have the lowest DE while the late-type
galaxies (Scd--Irr) have the highest DE, due to the presence of bright
nuclear regions in the early-type galaxies that tend to obscure
SNe. In fact, this becomes a serious issue for early-type galaxies
having small sizes. As shown in the inset of Figure 11, the early-type
galaxies with MaA smaller than $1'$ have rather poor DE even when the
detection is not limited by the brightness of the SNe. As discussed in
Papers II and III, we eliminated the small early-type galaxies and the
associated SNe in our final rate calculations by using the ``optimal''
subsamples.

The Monte Carlo simulations suggest that $\sim 10$\% of the SNe with
magnitudes much brighter than the limiting magnitude are missed in the
nuclear regions of the galaxies due to our SN search strategy, which
is consistent with the estimate from the radial distribution of the
SNe.  The detection efficiency takes into account this missed fraction, so
our final rates are not affected.

\section{Conclusions} 

This is Paper I of a series aiming to determine the rates of different
types of SNe in nearby galaxies, using data obtained from the Lick
Observatory Supernova Search during the past decade.  Here, we first
provide an outline of the series (\S 2.1), discuss the improvements of
our rate calculations over published results (\S 2.2), and explore
possible limitations (\S 2.3).

We then provide a mathematical derivation for the control-time
calculation for a SN type having a known luminosity function (LF)
represented by discrete components, with details in the Appendix.  Not
surprisingly, the total control time is the sum of the control times
for each SN weighted by its fraction in the LF.

We provide details on the construction of our total galaxy sample and
the different galaxy subsamples.  Although our total galaxy sample has
a strong deficit of low-luminosity galaxies, the galaxy sample within
60~Mpc is representative of a complete galaxy sample for $L_K \simgt 7
\times 10^{10}\,{\rm L}_\odot$.  Moreover, there is a sufficiently
large number of faint galaxies in our sample to provide good
statistics for the rate calculations (as demonstrated in Paper
III). We also discuss the properties of our sample galaxies and their
change over distance.  There is a strong Malmquist bias, so the
low-luminosity galaxies become increasingly incomplete, leading to a
monotonic increase in the average galaxy size with increasing
distance. This is an important trend that will be used in Paper III to
demonstrate a correlation between the SN rates and galaxy sizes.

The construction of the SN sample is discussed. In total, 1036 SNe
were found directly or independently by LOSS in the LOSS fields, and
929 were discovered in the galaxies considered for the rate
calculations. Several SN subsamples are constructed, an important one
of which excludes all SNe that occurred in small (MaA $< 1'$),
early-type galaxies and in highly inclined ($i > 75^\circ$) spiral
galaxies; it has 726 SNe and is the ``optimal'' subsample for the final
rate calculation.  We also confirm a significant difference between
the Hubble-type distributions of SNe~Ia and CC~SNe, as we had reported
previously (van den Bergh et al. 2002, 2003, 2005). A small number of
CC~SNe ($< 3$\%) were found in early-type galaxies, mostly in S0/a
galaxies that probably have low-level, recent star formation. The host
galaxies of the remaining CC~SNe show other signs of recent star
formation, or are misclassified spirals. Hence, CC~SNe in E--S0
galaxies are very rare, and this places a strong limit on the amount
of recent star formation in these galaxies. We also find that the
subclass of subluminous, peculiar, ``Ca-rich SNe~Ibc'' have a high
fraction in early-type galaxies, consistent with recent suggestions
that they come from low-mass stars in old populations (Perets et
al. 2010).

Details on the log files, and on how the limiting magnitude of each
image is calculated from information in the log files, are
provided. The limiting magnitude is a function of the intensity ratio
of the new and template images, the stellar FWHM, and the sky
background. Our limiting magnitude can be measured to a precision of
0.2--0.3 mag for any individual image in the survey.  We also show
that for the more than 2 million observations considered in this SN
rate calculation, the average observation interval is $\sim 9$~d,
which in most cases is much smaller than the control time for the
different types of SNe. Consequently, the contribution to the control
time is often the observation intervals themselves (multiplied by the
detection efficiency). This ensures that our control times have a
great degree of tolerance to uncertainties in the SN LFs.

We perform Monte Carlo simulations to study the detection efficiency
(DE) in our survey. Simulated SNe with different brightness are
injected into the images, which are then processed with the SN search
software used in our survey to study the fraction of the recovered
SNe. We find that the DE curves are different for galaxies of
different Hubble types, with the early-type galaxies having the lowest
DE and the late-type galaxies having the highest DE.  This is probably
caused by the higher missed fraction of SNe in the brighter nuclei of
the early-type galaxies. Overall, $\sim 10$\% of the injected bright
SNe are missed in the simulations due to our search strategy,
consistent with an estimate from a study of the radial distribution of
SNe. The DE curves take into account this missed fraction in the final
rate calculations.

It is demonstrated that the very nearby LOSS galaxies (within 60~Mpc)
are representative of galaxies bigger than the Milky Way, but include
many faint galaxies as well (though with an apparent deficit compared
to the complete sample).  There are abundant galaxies of different
Hubble types within 60~Mpc (Figures 2 and 4), and our survey may have
nearly full control over the different types of SNe in these galaxies
(Figure 6).  In Paper II, we will construct LFs for the CC~SNe using SNe
discovered in these nearby LOSS galaxies.  Because of their extreme
brightness, our survey is complete for SNe~Ia to a greater distance,
so the cutoff distance is set at 80~Mpc.

After deriving the limiting magnitude and the detection efficiency for
any individual image in our survey, we need two additional pieces of
information to calculate the control time for any given type of SN:
the distribution of the intrinsic brightness of SNe (i.e., the LF) and
their light-curve shapes.  These are the main results from Paper II,
where for the first time a complete SN sample is constructed and the
observed LFs are derived.

\section*{Acknowledgments}

We thank the referee, Enrico Cappellaro, for useful comments and
suggestions which improved the paper. We are grateful to the many
students, postdocs, and other collaborators who have contributed to
the Katzman Automatic Imaging Telescope and the Lick Observatory
Supernova Search over the past two decades, and to discussions
concerning the determination of supernova rates --- especially Ryan
J. Foley, Mohan Ganeshalingam, Saurabh W. Jha, Maryam Modjaz, Dovi
Poznanski, Frank J. D. Serduke, Jeffrey M. Silverman, Nathan Smith,
Thea Steele, Richard R. Treffers, and Xiaofeng Wang.  We also
acknowledge S. Bradley Cenko and Sydney van den Bergh for useful
discussions.  We thank the Lick Observatory staff for their assistance
with the operation of KAIT.

LOSS, conducted by A.V.F.'s group, has been supported by many grants
from the US National Science Foundation (NSF; most recently
AST-0607485 and AST-0908886), the TABASGO Foundation, US Department of
Energy SciDAC grant DE-FC02-06ER41453, and US Department of Energy
grant DE-FG02-08ER41563. KAIT and its ongoing operation were made
possible by donations from Sun Microsystems, Inc., the Hewlett-Packard
Company, AutoScope Corporation, Lick Observatory, the NSF, the
University of California, the Sylvia \& Jim Katzman Foundation, and
the TABASGO Foundation.  We give particular thanks to Russell
M. Genet, who made KAIT possible with his initial special gift; former
Lick Director Joseph S. Miller, who allowed KAIT to be placed at Lick
Observatory and provided staff support; and the TABASGO Foundation,
without which this work would not have been completed.  J.L. is
grateful for a fellowship from the NASA Postdoctoral Program. We made
use of the NASA/IPAC Extragalactic Database (NED), which is operated
by the Jet Propulsion Laboratory, California Institute of Technology,
under contract with NASA. We acknowledge use of the HyperLeda database
(http://leda.univ-lyon1.fr).

\newpage

\renewcommand{\baselinestretch}{1.0}

\newpage

\begin{figure*}
\includegraphics[scale=0.9,angle=270,trim=0 50 0 0]{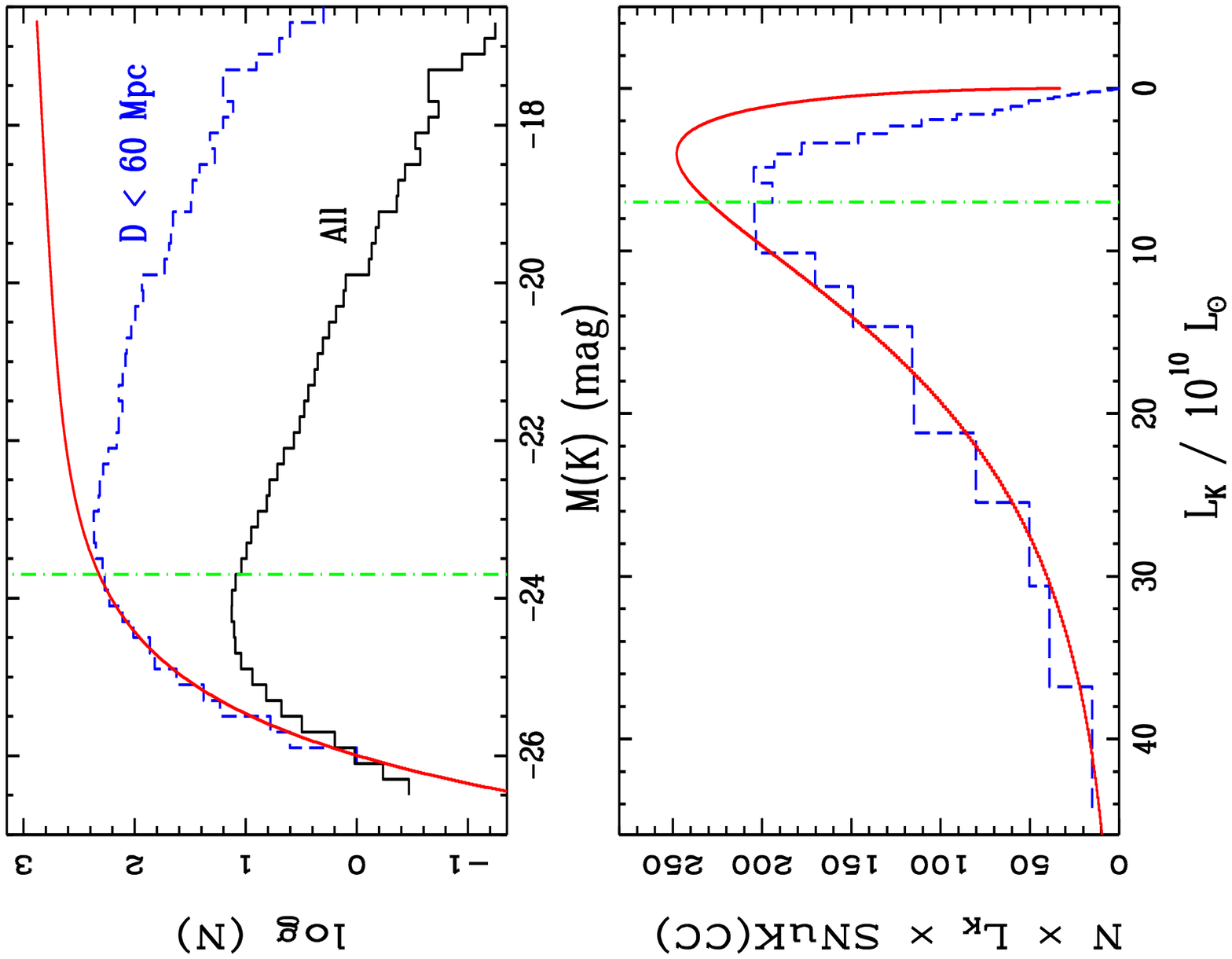} 
\caption{{\it Upper panel: } $K$-band luminosity function for all of
  the LOSS galaxies (solid steps) and for the LOSS galaxies within 60~Mpc 
  (dashed steps) versus the galaxy luminosity function for a
  complete sample (smooth curve; Kochanek et al. 2001). 
  {\it Lower panel: } The effective contribution to the number of CC~SNe
  for each luminosity bin. The curves represent the same samples as
  in the upper panel, except that the curve for the ``full" sample is not
  shown. In both panels, the dot-dashed line marks the nominal galaxy
  size ($L_K = 7 \times 10^{10}~ {\rm L}_\odot$) used for the final
  rate calculation in Paper III. }
\label{1}
\end{figure*}

\begin{figure*}
\includegraphics[scale=0.7,angle=270]{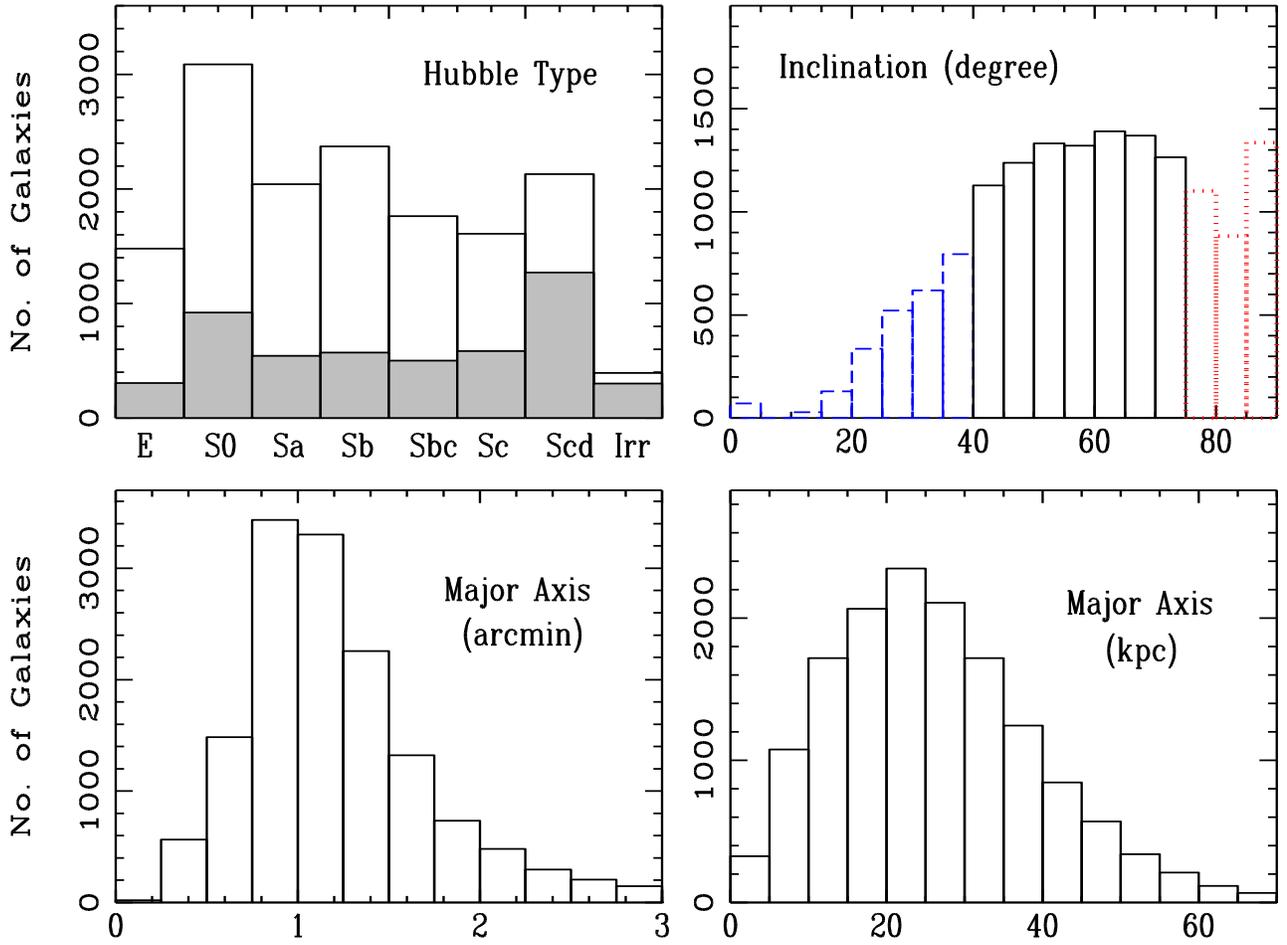} 

\caption[] {Some statistics regarding the LOSS galaxy sample. 
   {\it Upper left}: the Hubble-type distribution for all of the galaxies 
  (open histogram) and for galaxies within 60~Mpc (shaded histogram). 
   {\it Upper right}: the distribution of the inclination angles. The
  inclination angles are broadly categorised into three bins: face-on
  (dashed line), normal inclination (solid line), and edge-on (dotted
  line). {\it Lower left}: the size (MaA, in arcmin) distribution of the
  galaxies. {\it Lower right}: the size (MaA, in kpc) distribution of the
  galaxies.}

\label{2}
\end{figure*}

\begin{figure*}
\includegraphics[scale=0.9,angle=270,trim=0 100 0 0]{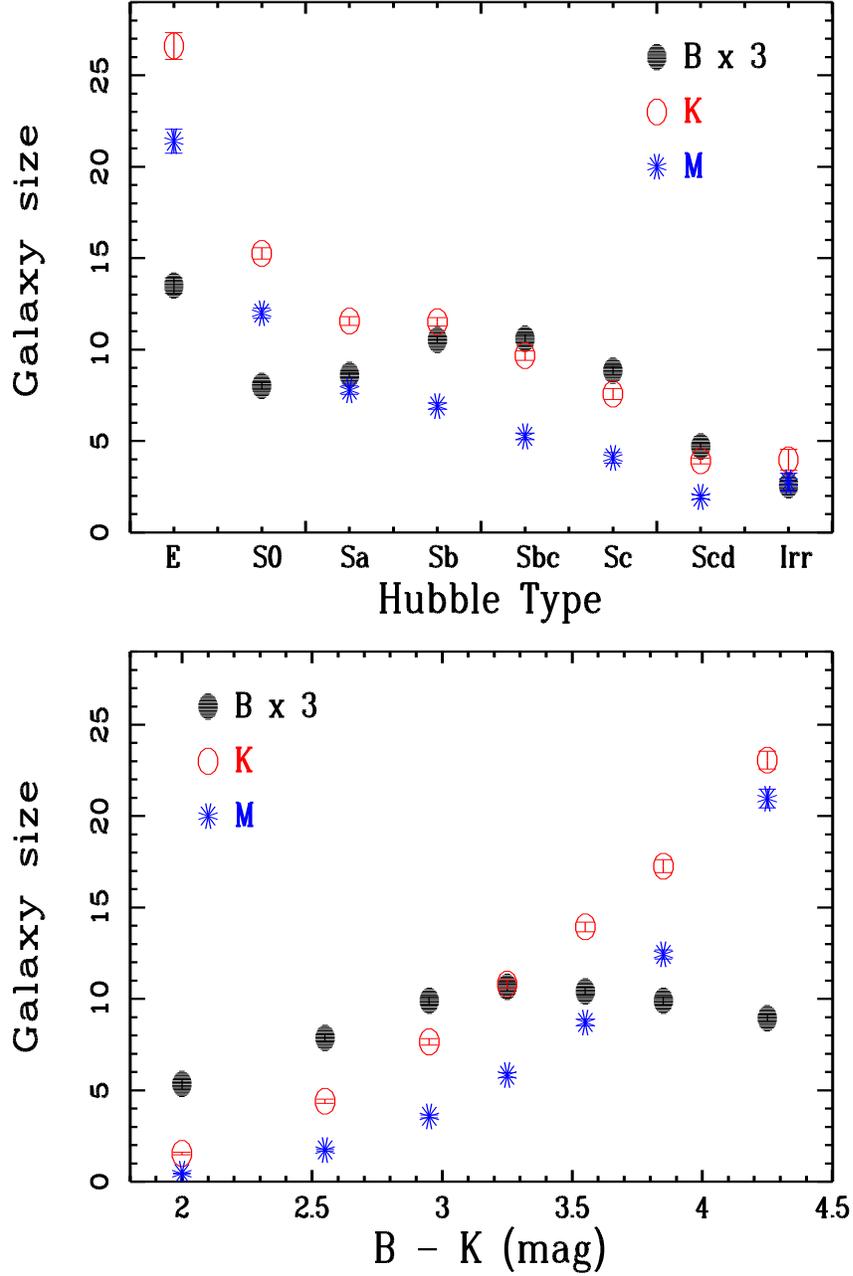} 
\caption[] {The average size of the galaxies versus galaxy Hubble
  type and $B - K$ colour. For the $B$ and $K$ luminosity, the size is
  in units of $10 \times 10^{10}~ {\rm L}_\odot$, while for the mass, the
  size is in units of $10 \times 10^{10}~ {\rm M}_\odot$.
}
\label{3}
\end{figure*}

\begin{figure*}
\includegraphics[scale=0.7,angle=270]{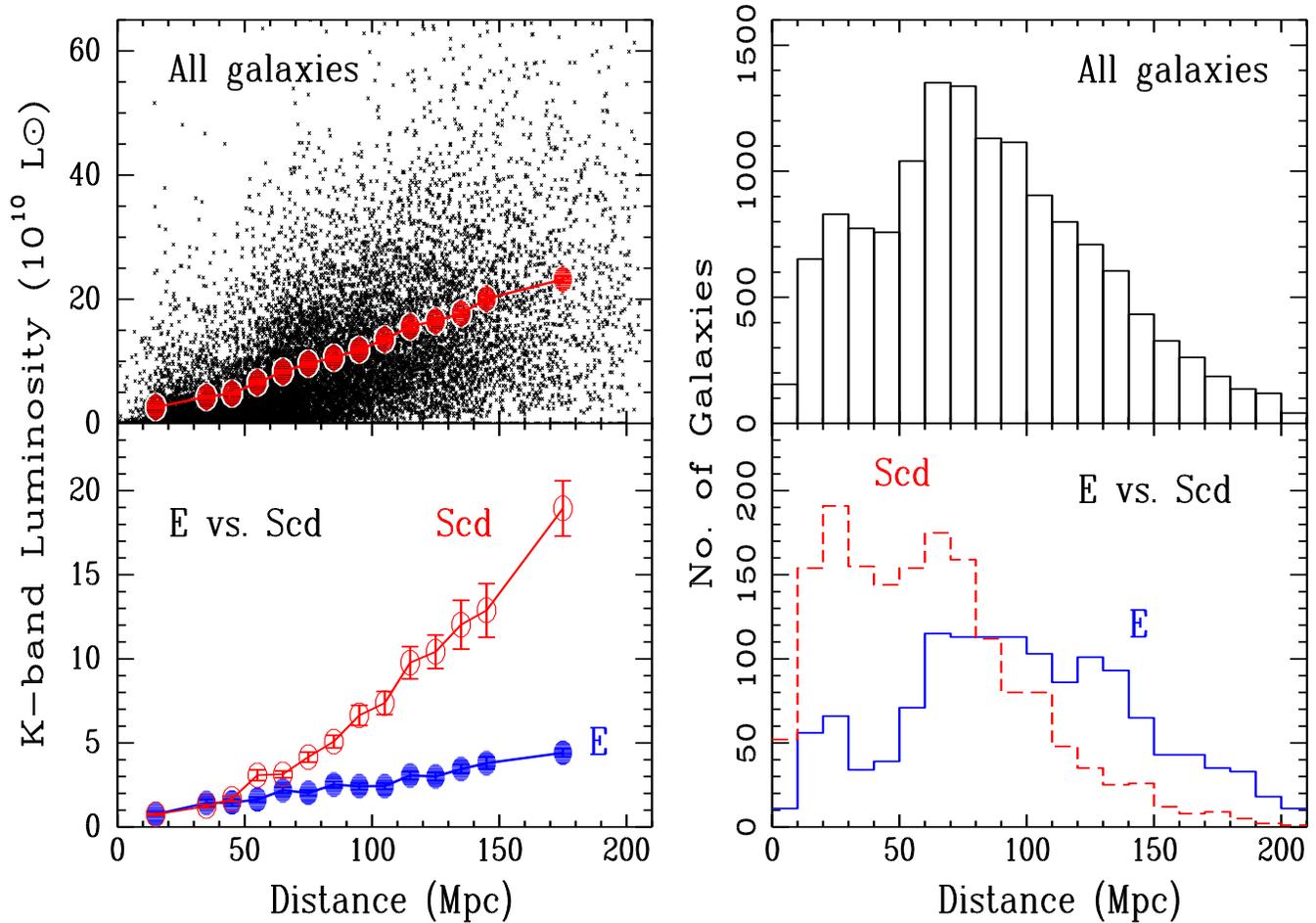} 
\caption[] {The change of the LOSS galaxy properties over distance.
  The upper-left panel shows the $K$-band luminosity. Each dot
  represents a galaxy, while the average luminosity in different
  distance bins is overplotted as big solid dots connected by a
  line. The lower-left panel shows the change of the average $K$-band
  luminosity over distance for two galaxy Hubble types. The Scd
  galaxies have a more dramatic change than do the E galaxies.  The
  upper-right panel shows the number distribution for all of the
  galaxies over distance, while the lower-right panel shows this for
  two different Hubble types. }
\label{4}
\end{figure*}

\begin{figure*}
\includegraphics[scale=0.8,angle=270]{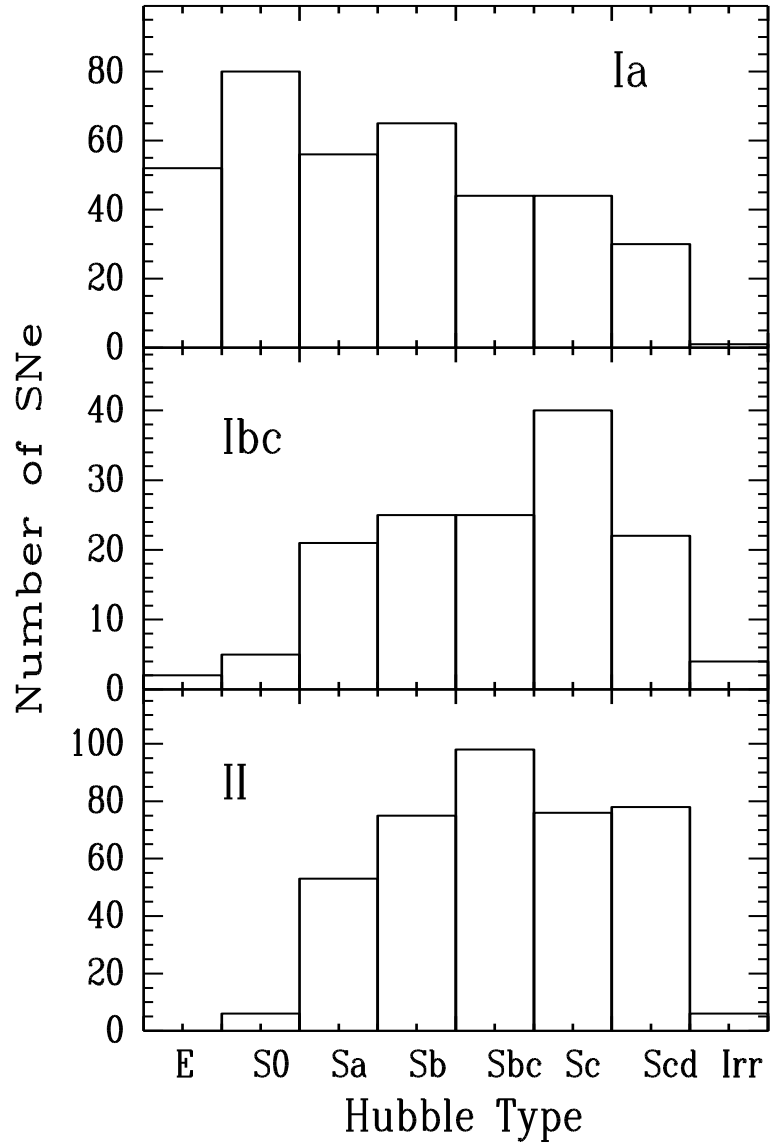} 
\caption[] {The Hubble-type distribution for the host galaxies of
  different types of SNe. There is a significant difference between
  the distribution of SN~Ia hosts and SN~Ibc/II hosts, while the 
  SN~Ibc and SN~II hosts have similar distributions.  }
\label{5}
\end{figure*}

\begin{figure*}
\includegraphics[scale=0.7,angle=270]{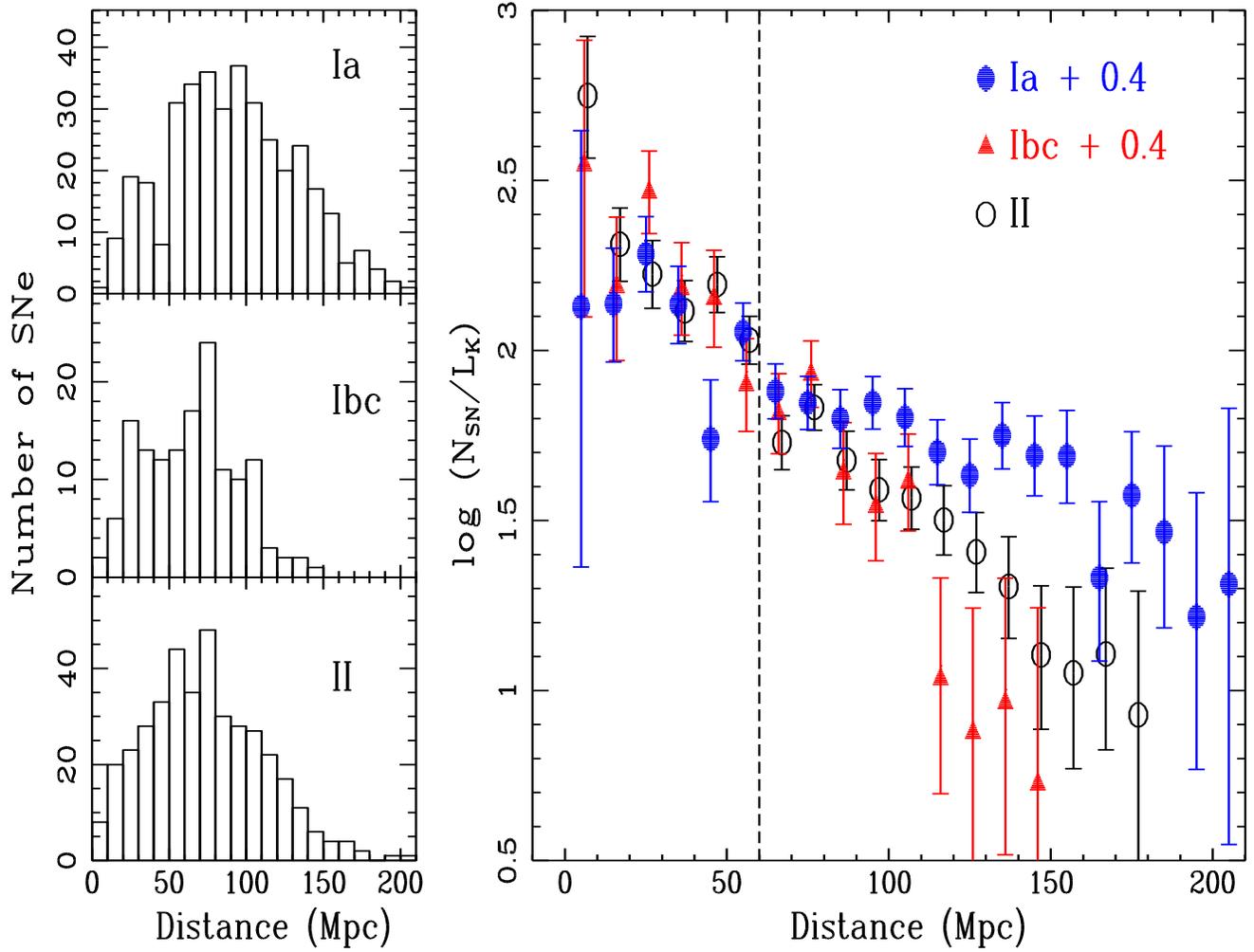} 
\caption[] {The left panels show the distribution of SNe over
  distance.  The right panel shows the ratio of the number of SNe in
  each bin divided by the total $K$-band luminosity for all the
  galaxies in the same distance bin.}
\label{6}
\end{figure*}

\begin{figure*}
\includegraphics[scale=0.9,angle=270,trim=0 70 0 0]{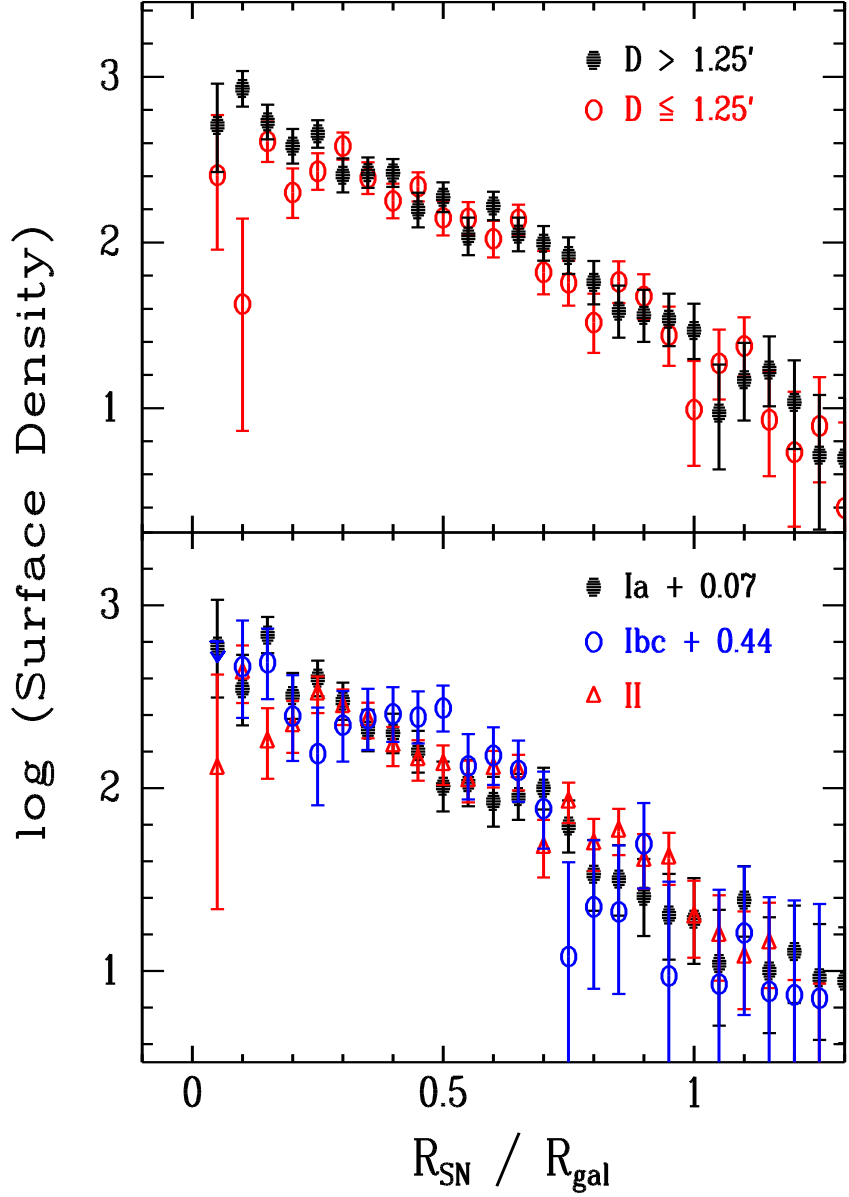} 
\caption[] { The radial distribution of SNe in their host
  galaxies.  The upper panel shows the distribution for two groups of
  SNe, one with relatively large host galaxies (MaA $> 1.25'$) and the
  other with smaller galaxies. The lower panel shows the radial 
  distribution for different types of SNe.}
\label{7}
\end{figure*}

\begin{figure*}
\includegraphics[scale=0.9,angle=270,trim=0 70 0 0]{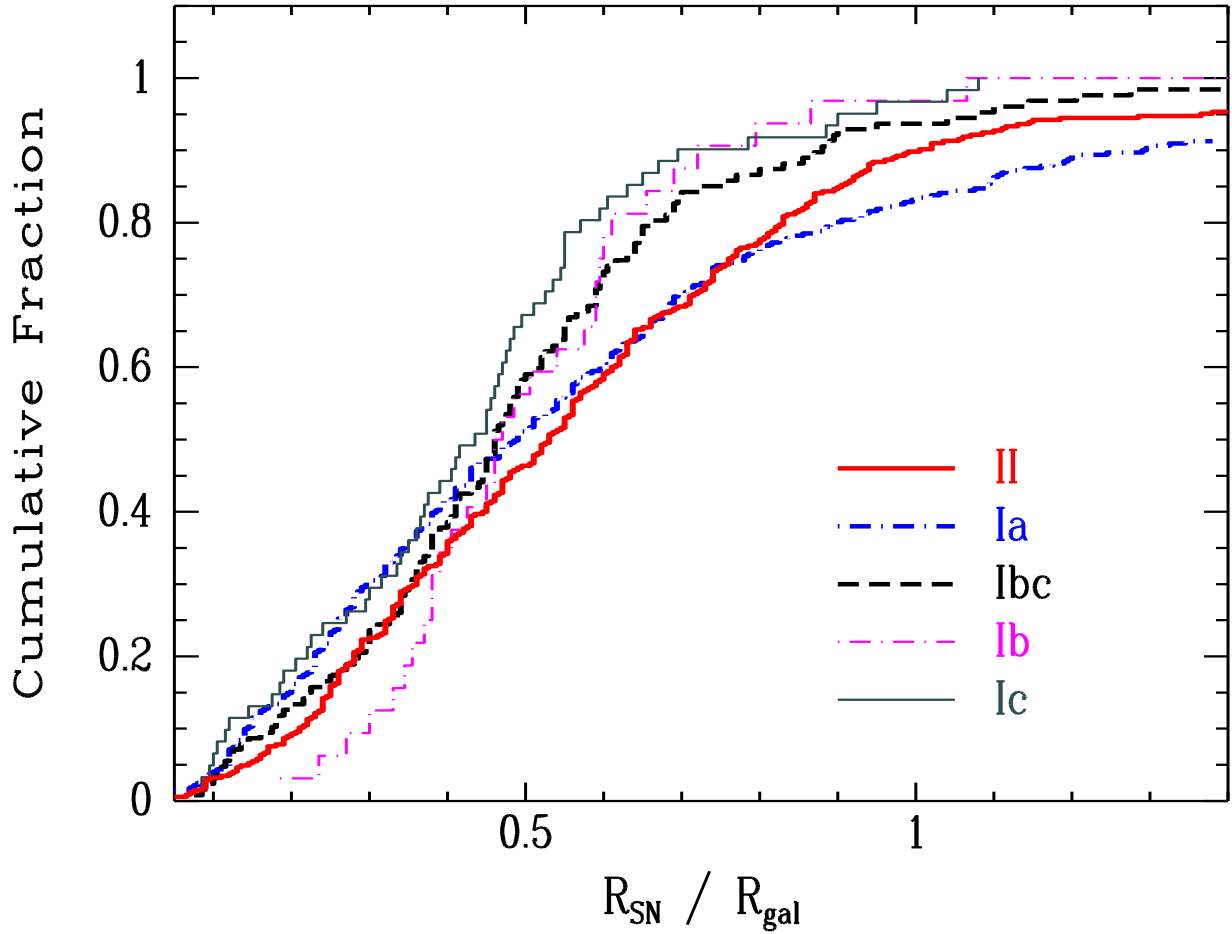} 
\caption[] {The cumulative fractions for the radial
  distributions of SNe~Ia (thick dash-dotted line), Ibc (thick dashed
  line), II (thick solid line), Ib (thin dash-dotted line), and Ic
  (thin solid line).  }
\label{8}
\end{figure*}

\begin{figure*}
\includegraphics[scale=0.8,angle=270]{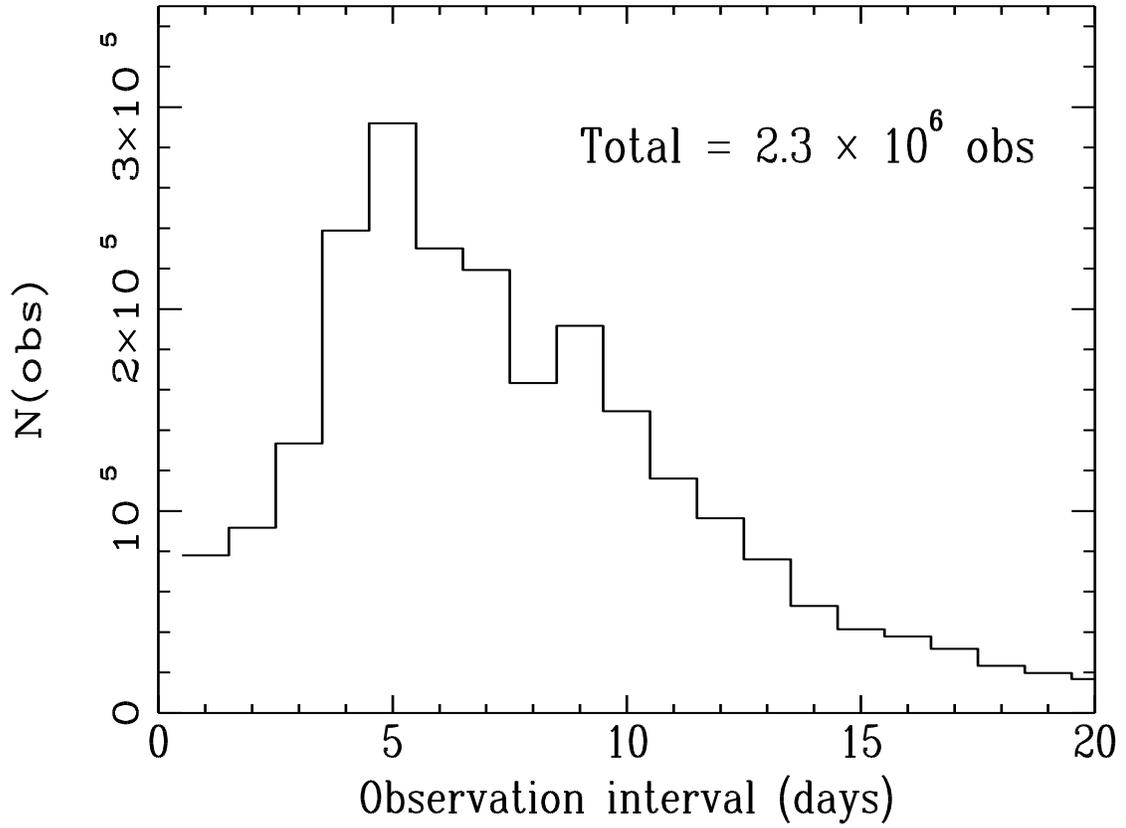} 
\caption[] { The distribution of the observation intervals for over 2
  million observations considered in the current SN rate calculation.  }
\label{9}
\end{figure*}

\begin{figure*}
\includegraphics[scale=1.0,angle=270,trim=0 100 100 0]{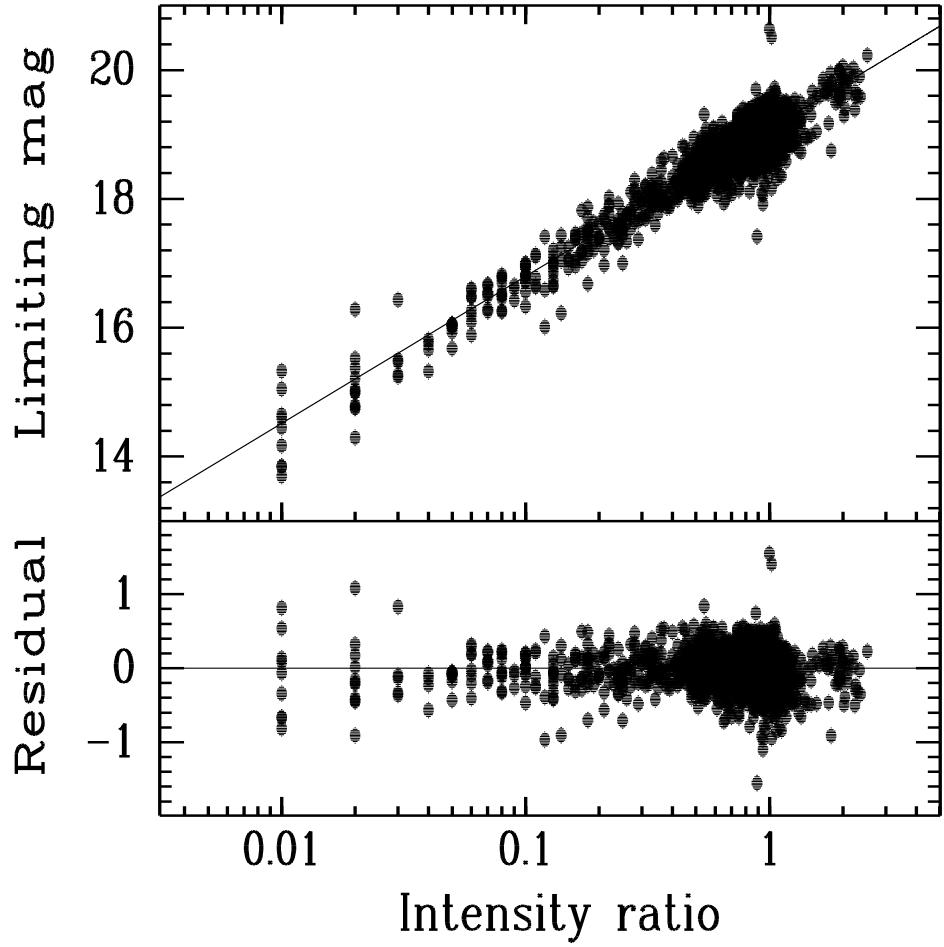} 
\caption[] {The correlation between the limiting magnitudes and the
  intensity ratio between the search image and the template. The limiting
  magnitudes have been corrected for the FWHM and sky background of
  the images to highlight the correlation shown here. }
\label{10}
\end{figure*}

\begin{figure*}
\includegraphics[scale=0.7,angle=270,trim=0 -30 0 0]{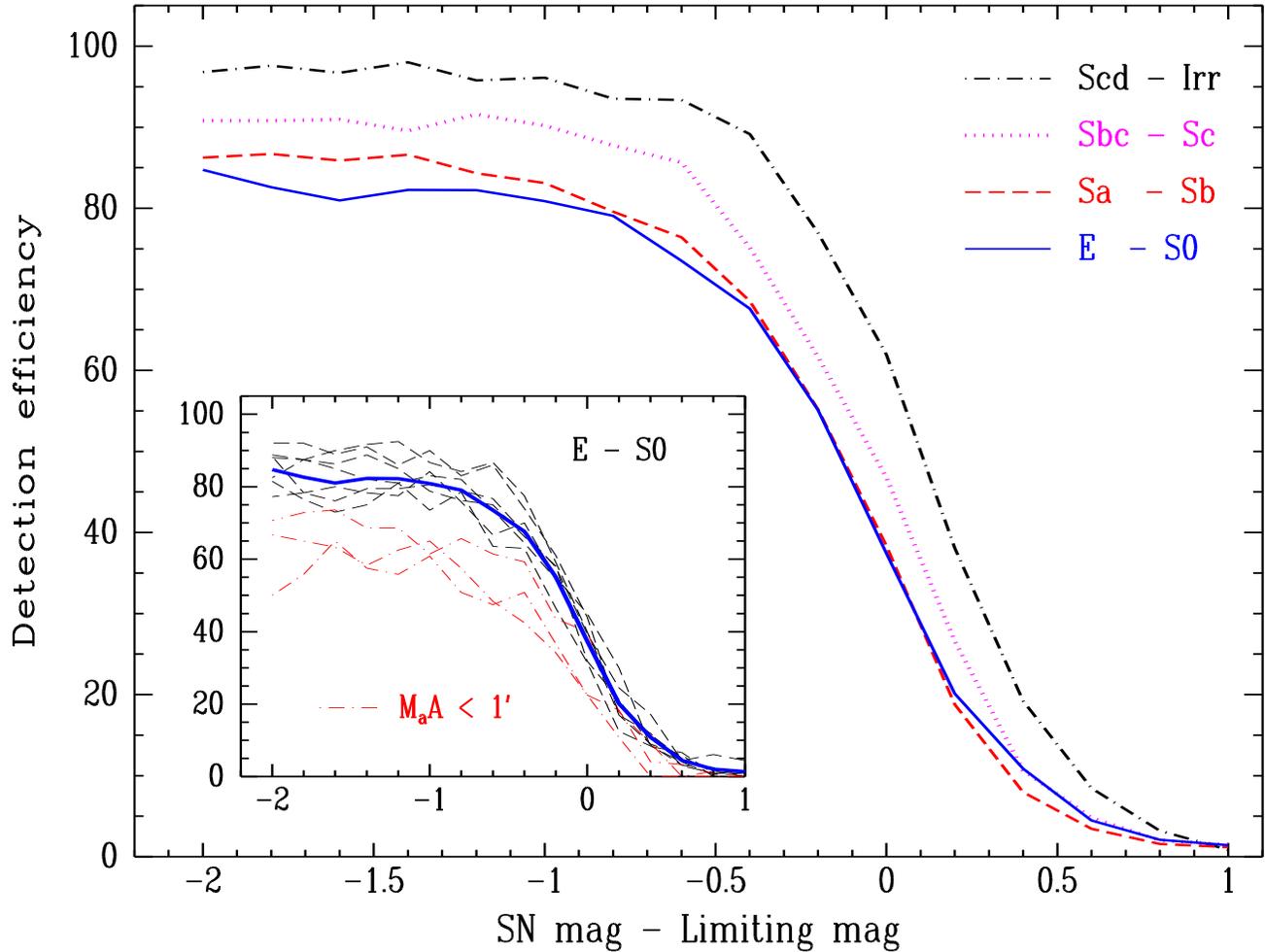} 
\caption[] { The detection efficiency (DE) in our survey. The DE
  curves have a strong dependence on the galaxy Hubble type, with the
  late-type galaxies having the highest DE and the early-type galaxies
  having the lowest DE. The inset shows the construction of the DE
  curve for the E--S0 galaxies. Each dashed line is the DE curve
  for one galaxy. The dash-dotted lines are for galaxies with small
  sizes (MaA $< 1'$), and they are not included in calculating
  the average DE curve (thick line). }
\label{11}
\end{figure*}

\vfil 
\clearpage

\scriptsize

\begin{table*}
\centering
\begin{minipage}{140mm}
\caption{Hubble-type definitions.}
\begin{tabular}{llll}
\hline
\hline
Number\footnote{The number sequence for the Hubble types used in this study.}
& Symbol\footnote{The symbols used in all of the figures.} & Type
& $T$\footnote{The range of $T$ values, a numerical code for the Hubble types
as defined by RC3.}\\
\hline
 1          &          E   &   E, E--S0    & $-5$ to $-3$  \\
 2          &          S0  &   S0, S0a    & $-2$ to 0   \\
 3          &          Sa  &   Sa, Sab    & 1 to 2\\
 4          &          Sb  &   Sb         & 3\\
 5          &          Sbc &   Sbc        & 4\\
 6          &          Sc  &   Sc         & 6\\
 7          &          Scd &   Scd, Sd, Sm& 7 to 9\\
 8          &          Irr &   Irr        & 10\\
\hline
\hline
\end{tabular}
\end{minipage}
\end{table*}

\begin{table*}
\vbox to330mm{\vfil Landscape Table 2 to go here.
\caption{Galaxy properties in the LOSS ``full" sample.}
\vfil}
\label{landtable} 
\end{table*}

\begin{table*}
\vbox to330mm{\vfil Landscape Table 3 to go here.
\caption{Average galaxy properties in the ``full'' and ``optimal'' samples.}
\vfil}
\label{landtable} 
\end{table*}

\begin{table*}
\vbox to330mm{\vfil Landscape Table 4 to go here.
\caption{The supernova sample.}
\vfil}
\label{landtable} 
\end{table*}

\begin{table*}
\caption{Core-collapse SNe in early-type galaxies.}
\begin{tabular}{lllclcccccc}
\hline
\hline
{SN} &{Type} & {Host} &{NED} &{HyperLeda} &{H2008} 
&{Comment} \\
\hline
 1999ew &II  &NGC-3677     & S0/a     & S0/a    & ...& \\
 2000ds &Ibcp&NGC-2768     & S0/a     & E       & S0       &``Ca-rich"    \\
 2002aq &II  &MCG-01-7-35  & S0       & S0/a    & ...& SN in a star-forming ring \\
 2003ei &IIn &UGC-10402    & ...& E       & ...& SN in a tidal arm \\
 2004gh &II  &MCG-04-25-6  & S0/a     & S0/a    & ...& \\
 2005E  &Ibcp&NGC-1032     & S0/a     & S0/a    & ...& ``Ca-rich" \\
 2005ar &Ib  &CGCG-011-033 & E        & E       & ...& Possible S0/a galaxy \\
 2005lw &II  &IC-672       & ...& S0/a    & ...& \\
 2006ee &II  &NGC-774      & S0       & S0      & S0       & \\
 2006lc &Ib  &NGC-7364     & S0/a     & S0/a    & ...& \\
 2007aw &Ic  &NGC-3072     & S0/a?    & S0/a    & ...& \\
 2007ke &Ibcp&NGC-1129     & E        & E       & ...& In a cluster; ``Ca-rich"  \\
 2007kj &Ibc &NGC-7803     & S0/a     & S0/a    & ...& \\
\hline
\hline
\end{tabular}
\end{table*} 

\vfil 
\clearpage

\appendix

\section{Details of the Control-Time Method}

The control-time method is used in our rate calculations. Here we
first describe how the method is implemented in the most basic
scenario, with a single light-curve shape and luminosity for a given
type of SN.  We then describe how the rates are calculated when the SN
type has a range of luminosities and light-curve shapes --- that is,
with a known SN luminosity function (LF).

\subsection{The Control-Time Method Using a Single Light Curve} 

Although the method in the most basic scenario is well documented in
the literature (e.g., Zwicky 1942; van den Bergh 1991; Cappellaro et
al. 1993a, 1997), here we include a brief description
for completeness. We follow the methodology used by Cappellaro et
al. (1997) with some minor modifications.

Let $t_1, t_2, ..., t_i, ... $ be the epochs of observations, so $t_i
- t_{i-1}$ is the time interval between observations $i-1$ and
$i$. For a possible SN in the $j$-th galaxy, the control time for a
single image at epoch $t_i$, $C_{i,j}$, can be evaluated by
calculating the interval of time during which the SN stays brighter
than the limiting magnitude of the observation.  Obviously, $C_{i,j}$
depends on the adopted peak magnitude of the SN, the light-curve shape
of the SN, and the limiting magnitude of the observation.

The total control time for this $j$-th galaxy with a total of $n$
observations, $tC_j$, can be computed as
\begin{equation}
tC_j = \sum_{i=1}^{n} \Delta t_i  c_i,
\end{equation}
\noindent
where
\begin{equation}
\Delta t_i   = 
\left\{
\begin{array}{ll}
C_{i,j} & if\,\, t_i - t_{i-1} \ge C_{i,j}\,\, {\rm or}\,\, i = 1 \\
t_i - t_{i-1} & {\rm otherwise},
\end{array}
\right.
\end{equation}
\noindent
and $c_i$ is a correction factor introduced to account for the bias
against SN discovery in the nuclear regions of galaxies in the
historical rate calculations. In our study, we perform Monte Carlo
simulations (\S 4.5) to account for the detection efficiency of each
observation, rather than adopting a global correction factor.

We further define the total normalised control time as 
\begin{equation}
t_j = tC_jL_j,
\end{equation}
\noindent
where $L_j$ is the galaxy luminosity (or mass).
Finally, the SN rate is calculated for a galaxy sample with 
$N_{\rm G}$ galaxies and $N_{\rm SN}$ SNe as
\begin{equation}
{\rm rate} = \frac{N_{\rm SN}}{\sum_{j=1}^{N_G} t_j}.
\end{equation}

In our survey, the galaxies in the sample were observed with a
relatively short time interval (\S 4.3), so the vast majority of
control times are derived solely from the observation intervals
(multiplied by the detection efficiency).  Because of this, our rates
have a great degree of tolerance for differences in the adopted SN
light-curve shapes and LFs, especially for SNe~Ia, the most luminous
class. However, there are still instances where the control time needs
to be derived from the limiting magnitude and the light curve, such as
after a long interval of bad weather or, especially, when the galaxy
under consideration reemerges into the nighttime sky.

\subsection{The Control-Time Method using a Known Luminosity Function}

SNe display a great degree of diversity in their peak absolute
luminosity and their photometric behaviours (e.g., Leibundgut et
al. 1991; Filippenko 1997; Richardson et al. 2002). Even for the most
homogeneous class, SNe~Ia, a large fraction are either the
fast-declining subluminous SN 1991bg-like objects or the slow-evolving
SN 1991T-like events (Li et al. 2001). Consequently, treating a given
SN type as having a single light-curve shape with a single peak
absolute magnitude is an oversimplification, and has the potential to
introduce large uncertainties in the final rates. In previous studies,
this problem has been partly dealt with by adopting a Gaussian scatter
to the peak absolute magnitudes and sometimes stretching the light
curve of a SN~Ia according to its luminosity (e.g., Cappellaro et
al. 1997; C99; Barris et al. 2006; Neill et al. 2006; Sullivan et
al. 2006 Poznanski et al. 2007; Botticella et al. 2008; Sharon et
al. 2008).

As we report in Paper II, however, a Gaussian scatter is not a good
approximation to the LFs of the SNe. Instead, our LFs consist of
discrete peak absolute magnitudes from a complete sample of SNe, with
a family of light curves for each type (component) of SN. In this
section, we show how the control time is calculated for a SN type with
a known LF. Not surprisingly, the final control time is the sum of the
control time for each component weighted by its relative fraction in
the LF.

Let us consider a SN LF with $n$ components and relative fractions of
$f_1, f_2, ..., f_n$. We examine two scenarios. In the
first scenario, a survey has complete control of every component of
the SN LF during a total normalised control time of $t$, and a yield of
$N$ discoveries.  If we use $t_i$ as the total normalised control
time and $N_i$ as the number of SNe for the $i$-th component of the
LF, we have the following equations:
\begin{equation}
\begin{array}{ll}
\sum_{i=1}^{n} f_i = 1.0  \\
\\
\vspace*{0.1 truein}
t_i = t & i = 1, 2, ... n \\
\vspace*{0.1 truein}
N_i = f_i \times N & i = 1, 2, ... n \\
\vspace*{0.1 truein}
{\rm rate} = \sum_{i=1}^{n}\frac{N_i}{t_i} = \frac{N}{t}.
\end{array}
\end{equation}

In the second scenario, a survey has partial control of the individual
components. For the $i$-th component, $N_i^\prime$ SNe are discovered
with a total normalised control time of $t_i^\prime$.
Under this assumption, $t_i^\prime \leq t$.

Comparing the second scenario to the first, one has the following 
equations according to the concept of control times:
\begin{equation}
N_i^\prime / t_i^\prime = N_i / t_i = N_i / t
\end{equation}
\noindent
This equation can be rewritten as
\begin{equation}
N_i^\prime  =\frac{N_i}{t} \times t_i^\prime.
\end{equation}
\noindent
The rate of the second scenario can be calculated as
\begin{equation}
\begin{array}{lcl}
{\rm rate} & = & \frac{N_1^\prime}{t_1^\prime} +
\frac{N_2^\prime}{t_2^\prime} + ... + \frac{N_n^\prime}{t_n^\prime}
\\ & = & (N_1^\prime + N_2^\prime + ... + N_n^\prime) \times \\ & &
   [ \frac{N_1^\prime}{(N_1^\prime + N_2^\prime + ... +
       N_n^\prime)} \times \frac{1}{t_1^\prime} \\ & & +
     \frac{N_2^\prime}{(N_1^\prime + N_2^\prime + ... +
       N_n^\prime)} \times \frac{1}{t_2^\prime} \\ & & + ... +
     \frac{N_n^\prime}{(N_1^\prime + N_2^\prime + ... +
       N_n^\prime)} \times \frac{1}{t_n^\prime} ].

\end{array}
\end{equation}

We note that $(N_1^\prime + N_2^\prime + ... + N_n^\prime) =
N^\prime$ is the total number of observed SNe in the second
scenario. Substituting Equation (A7) into the $i$-th component within
the parentheses of Equation (A8) yields the following:
\begin{equation}
\begin{array}{ccl}
\frac{N_i^\prime}{(N_1^\prime + N_2^\prime + ... + N_n^\prime)}
\times \frac{1}{t_i^\prime} & = & \frac{\frac{N_i}{t}
  t_i^\prime}{\frac{N_1}{t} t_1^\prime + \frac{N_2}{t} t_2^\prime +
  ... + \frac{N_n}{t} t_n^\prime} \times \frac{1}{t_i^\prime} \\ & =
& \frac{N_i}{N_i t_1^\prime + N_2 t_2^\prime + ... + N_n
  t_n^\prime}.
\end{array}
\end{equation}
\noindent
When Equation (A9) is substituted into Equation (A8), we have
\begin{equation}
\begin{array}{ccl}
{\rm rate} & = & N^\prime \times [ \frac{N_1}{N_1 t_1^\prime + N_2
    t_2^\prime + ... + N_n t_n^\prime} \\ & & + \frac{N_2}{N_1
    t_1^\prime + N_2 t_2^\prime + ... + N_n t_n^\prime} + ... \\ & &
  + \frac{N_n}{N_1 t_1^\prime + N_2 t_2^\prime + ... + N_n
    t_n^\prime} ] \\ & = & N^\prime \times [ \frac {N_1 + N_2 +
    ... + N_n}{N_1 t_1^\prime + N_2 t_2^\prime + ... + N_n^\prime
    t_n^\prime}] \\ & = & N^\prime \times [ \frac {N}{N_1
    t_1^\prime + N_2 t_2^\prime + ... + N_n^\prime t_n^\prime} ]
\\ & = & N^\prime \times [ \frac {1} {\frac{N_1}{N} t_1^\prime +
    \frac{N_2}{N} t_2^\prime + ... + \frac{N_n}{N} t_n^\prime} ]
\\ & = & \frac{N^\prime} {\sum_{i=1}^{n} f_i t_i^\prime} \\ & = &
\frac{N^\prime} {t^\prime},
\end{array}
\end{equation}
\noindent
where $t^\prime$ is the total control time for the survey. From Equation 
(A10), we obtain
\begin{equation}
t^\prime = \sum_{i=1}^{n} f_i t_i^\prime,
\end{equation} 
\noindent
which means that the total control time is the sum of the control time
of each component weighted by its fraction in the luminosity
function. Equation (A11) provides the foundation on how our control
time is calculated in Paper III.

\end{document}


\begin{landscape}

\setcounter{table}{1}
\begin{table*}
\begin{minipage}{295mm}
\caption{Galaxy properties in the LOSS ``full" sample.$^{a}$}
\begin{tabular}{lllrrlllllllrlllllll}
\hline
\hline
{field}&{host galaxy}&{$\alpha^\circ$ (J2000)}&{$\delta^\circ$}&{Dist$^{b}$}&
{src1$^{b}$}&{$h^{c}$}&{src2$^{c}$}&{$d1^{d}$}&{$d2^{d}$}&
{incl.$^{d}$}&{src3$^{d}$}&{PA$^{e}$}&{$E_{BV}^{f}$}&{$A_i^{g}$}&
{$B$}&{$B$(err)}&{$B0$}&{$K$}&{$K$(err)}
\\ 
\hline
{(1)} & {(2)} & {(3)} & {(4)}
&{(5)} &{(6)} &{(7)} &{(8)}
&{(9)} &{(10)} &{(11)} &{(12)}
&{(13)} &{(14)} &{(15)} &{(16)}
&{(17)} &{(18)} &{(19)} &{(20)}  \\
\hline
\hline
A2357+47  & UGC 12889           &   0.00705 &  47.27450 &  71.4 & LEDA &  4 & NED  &   1.87 &   1.18 &  55.4 & LEDA & 163.00 &   0.121 &   0.296 & 14.167 &  0.645 & 13.307 &  9.542 & 0.039\\
F00943    & UGC 12890           &   0.02925 &   8.27911 & 159.2 & NED  &  1 & NED  &   1.55 &   0.83 &  62.6 & LEDA &   9.50 &   0.060 &   0.000 & 15.675 &  0.780 & 15.417 & 10.625 & 0.051\\
F00984    & MCG -01-01-016      &   0.03600 &  -6.37397 &  89.0 & LEDA &  3 & NED  &   1.42 &   0.31 &  87.8 & LEDA & 167.75 &   0.045 &   0.531 & 14.843 &  0.381 & 14.033 & 11.125 & 0.055\\
F00982    & VIII Zw 494 NOTES01 &   0.08955 &  -2.61203 & 155.2 & LEDA &  1 & LEDA &   1.39 &   0.71 &  64.6 & LEDA & 127.51 &   0.038 &   0.000 & 15.170 &  0.422 & 14.834 & 10.950 & 0.062\\
UGC12893  & UGC 12893           &   0.11760 &  17.21952 &  16.0 & LEDA &  7 & NED  &   1.15 &   1.03 &  30.3 & LEDA &  95.00 &   0.034 &   0.089 & 15.383 &  0.390 & 15.141 & 13.452 & 0.223\\
F00473    & UGC 12897           &   0.15810 &  28.38447 & 120.9 & LEDA &  3 & NED  &   1.16 &   0.37 &  78.0 & LEDA &   7.20 &   0.062 &   0.543 & 14.822 &  0.075 & 13.925 & 10.369 & 0.040\\
F00472    & 2MASX J00004242+282 &   0.17670 &  28.36892 & 125.0 & LEDA &  2 & LEDA &   0.28 &   0.21 &  45.5 & LEDA &  94.91 &   0.062 &   0.000 & 17.716 &  0.143 & 17.314 & 12.965 & 0.155\\
F00472    & CGCG 498-061        &   0.18345 &  28.40142 & 118.5 & LEDA &  6 & NED  &   0.76 &   0.54 &  48.4 & LEDA &  87.30 &   0.062 &   0.135 & 15.661 &  0.119 & 15.150 & 11.367 & 0.066\\
F00472    & 2MASX J00004507+282 &   0.18780 &  28.37169 & 120.0 & LEDA &  2 & LEDA &   0.41 &   0.26 &  54.9 & LEDA & 158.83 &   0.062 &   0.081 & 17.258 &  0.168 & 16.781 & 13.107 & 0.162\\
F00472    & UGC 12899           &   0.19560 &  28.40197 & 121.7 & LEDA &  1 & NED  &   0.92 &   0.85 &  26.5 & LEDA & \nodat &   0.062 &   0.000 & 14.636 &  0.128 & 14.235 & 10.176 & 0.051\\
F00998    & ESO 538- G 017      &   0.23070 & -18.95911 & 103.7 & LEDA &  6 & NED  &   0.77 &   0.46 &  58.5 & LEDA &  27.22 &   0.021 &   0.352 & 15.569 &  0.349 & 15.071 & 13.004 & 0.139\\
F00460    & UGC 12900           &   0.23355 &  20.33792 &  94.3 & LEDA &  6 & NED  &   1.80 &   0.26 &  90.0 & LEDA & 110.00 &   0.078 &   1.413 & 15.611 &  0.329 & 13.823 & 11.151 & 0.061\\
UGC12901  & UGC 12901           &   0.24540 &  28.91169 &  96.1 & LEDA &  4 & NED  &   1.26 &   0.50 &  72.3 & LEDA &  44.50 &   0.049 &   0.563 & 14.819 &  0.348 & 13.992 & 10.835 & 0.059\\
F00937    & NGC 7802            &   0.25200 &   6.24290 &  73.1 & LEDA &  2 & NED  &   1.18 &   0.80 &  51.8 & LEDA &  54.00 &   0.054 &   0.000 & 14.668 &  0.330 & 14.353 &  9.960 & 0.032\\
F00484    & UGC 12904 NOTES02   &   0.26520 &  34.65350 & 175.8 & LEDA &  3 & NED  &   0.84 &   0.82 &  14.9 & LEDA & \nodat &   0.073 &   0.007 & 15.461 &  0.383 & 14.946 & 11.068 & 0.064\\
F00935    & IC 5374             &   0.26850 &   4.50001 & 122.5 & LEDA &  5 & LEDA &   0.68 &   0.58 &  35.6 & LEDA &  15.89 &   0.030 &   0.119 & 15.531 &  0.377 & 15.226 & 11.822 & 0.087\\
F00935    & IC 5375             &   0.27045 &   4.54200 & 125.4 & LEDA &  2 & NED  &   0.90 &   0.51 &  60.6 & LEDA & 177.00 &   0.030 &   0.388 & 15.313 &  0.334 & 14.726 & 11.040 & 0.047\\
F00950    & KUG 2358+128A       &   0.30585 &  13.14406 &  75.4 & LEDA &  6 & NED  &   0.75 &   0.30 &  72.5 & LEDA &  66.66 &   0.082 &   0.646 & 15.884 &  0.081 & 14.850 & 12.443 & 0.116\\
F00484    & UGC 12904           &   0.30900 &  34.67572 & 179.4 & LEDA &  3 & NED  &   1.02 &   0.69 &  52.3 & LEDA &  32.43 &   0.072 &   0.218 & 15.198 &  0.346 & 14.538 & 10.807 & 0.061\\
F00950    & MCG +02-01-010      &   0.31290 &  13.11330 &  77.2 & LEDA &  7 & NED  &   0.51 &   0.19 &  74.7 & LEDA &  50.00 &   0.081 &   0.725 & 16.773 &  0.202 & 15.663 & 14.140 & 0.264\\
\hline
\hline
\end{tabular}

\medskip

$^{a}$Only the first 20 entries are shown; the rest of the data
are available electronically. See text for details concerning the
meanings of the columns.

$^{b}$Distance (in Mpc) and source of the distance. NED = 
the NASA/IPAC Extragalactic Database (http://nedwww.ipac.caltech.edu);
LEDA = the HyperLeda Database (http://leda.univ-lyon1.fr).

$^{c}$The Hubble type (with the numbering scheme in Table 1) and 
the source of the Hubble type. 

$^{d}$The major diameter ($d1$, in arcmin), minor diameter ($d2$, 
in arcmin), inclination (incl., in degrees), and the source
of these data.

$^{e}$The position angle, in degrees.

$^{f}$The Galactic reddening $E(B-V)$, adopted from Schlegel,
Finkbeiner, \& Davis (1998).

$^{g}$The internal extinction due to inclination, in magnitudes.

\end{minipage}
\end{table*}

\clearpage
\newpage

\begin{table*} 
\begin{minipage}{260mm}
\caption{Average galaxy properties in the ``full'' and ``optimal'' samples.$^{a}$}
\begin{tabular}{@{}lrrrrrrrrrrrrr}
\hline
\hline
 & & & {Full} & &&&&&&{Optimal} &&&\\
\hline
\hline
{Hubble} &{$L(B)$} &{$L(K)$} &{$M$}
&{$N(B)$} &{$N(K)$} &{$N(M)$} & 
&{$L(B)$} &{$L(K)$} &{$M$}
&{$N(B)$} &{$N(K)$} &{$N(M)$} \\
\hline
E&3.30(09)&19.72(49)&15.98(44)&1445&1362&1327&\,\,\,\,\,\,\,\,\,\,\,\,\,\,\,\,\,&4.50(14)&26.61(74)&21.40(66)&750&729&724\\
S0&2.39(04)&13.32(22)&10.42(19)&3054&2989&2955&&2.67(06)&15.26(31)&11.99(27)&1915&1881&1875\\
Sa&2.95(05)&11.52(20)&7.59(19)&2027&1999&1982&&2.87(05)&11.56(24)&7.78(23)&1537&1513&1501\\
Sb&3.57(06)&11.00(19)&6.40(14)&2345&2302&2275&&3.51(06)&11.52(23)&6.91(18)&1657&1618&1604\\
Sbc&3.53(07)&9.17(21)&4.84(13)&1735&1682&1653&&3.53(08)&9.67(26)&5.25(17)&1220&1178&1162\\
Sc&2.96(07)&7.26(24)&3.80(22)&1587&1499&1477&&2.95(08)&7.57(30)&4.08(29)&1165&1102&1093\\
Scd&1.69(05)&3.79(13)&1.80(08)&2101&1627&1598&&1.57(05)&3.91(16)&1.94(10)&1472&1139&1120\\
Irr&0.93(08)&3.67(48)&2.41(40)&390&205&200&&0.86(09)&3.97(58)&2.74(49)&319&161&156\\
\hline
{$B-K$ (mag)} &{$L(B)$} &{$L(K)$} &{$M$}
&{$N(B)$} &{$N(K)$} &{$N(M)$} & 
&{$L(B)$} &{$L(K)$} &{$M$}
&{$N(B)$} &{$N(K)$} &{$N(M)$} \\
\hline
$<$2.3&1.99(08)&1.74(07)&0.50(02)&1703&1703&1703&&1.78(09)&1.54(07)&0.44(02)&1030&1030&1030\\
2.3$-$2.8&2.81(06)&4.71(10)&1.85(04)&1779&1779&1779&&2.62(07)&4.40(12)&1.73(05)&1200&1200&1200\\
2.8$-$3.1&3.48(07)&8.09(16)&3.78(07)&1774&1776&1774&&3.30(07)&7.65(18)&3.57(08)&1246&1248&1246\\
3.1$-$3.4&3.47(06)&10.55(19)&5.69(10)&2015&2015&2015&&3.55(08)&10.81(24)&5.83(13)&1435&1435&1435\\
3.4$-$3.7&3.21(06)&12.87(22)&8.03(14)&2144&2144&2144&&3.47(07)&13.94(27)&8.71(17)&1547&1547&1547\\
3.7$-$4.0&2.98(05)&15.59(27)&11.23(20)&2147&2147&2147&&3.30(07)&17.26(35)&12.43(25)&1475&1475&1475\\
$>$4.0&2.73(05)&20.86(37)&18.82(39)&1855&1855&1855&&2.98(06)&23.06(48)&20.96(51)&1276&1276&1276\\
\hline
\hline
\end{tabular}

\medskip

$^{a}$$B$ and $K$ luminosities are in units of $10^{10}\,{\rm
  L}_\odot$; mass is in units of $10^{10}\,{\rm M}_\odot$.

\end{minipage}
\end{table*}

\clearpage

\begin{table*}                                                                  
\begin{minipage}{295mm}                                                         
\caption{The supernova sample.$^{a}$}
\begin{tabular}{@{}llcrrccllccccccc}
\hline                                                                          
\hline                                                                          
SN & host galaxy & Disc. Date$^{b}$ & $\alpha^\circ$(J2000)&$\delta^\circ$
&Offset$^{c}$ &Mag$^{d}$ &Type &Discoverer
&full$^{e}$ &full-nos$^{e}$ &full-opt$^{e}$
&season$^{e}$&season-nos$^{e}$&season-opt$^{e}$
&LF
\\                                                                              
\hline                                                                          
{(1)} & {(2)} & {(3)} & {(4)}                                                   
&{(5)} &{(6)} &{(7)} &{(8)}                                                     
&{(9)} &{(10)} &{(11)} &{(12)}                                                  
&{(13)} &{(14)} &{(15)} &{(16)}\\                                              
\hline                                                                          
\hline                                                                          
1998S  & NGC 3877            & 1998-03-03 & 176.52574 &  47.48208 &  16.0W  46.0S & 15.2 & IIn     & BAOSS      & $\surd$ & $\surd$ &         &         &         &         &        \\
1998cu & IC 1525             & 1998-06-30 & 359.80396 &  46.87550 &  29.2W  50.3S & 18.1 & II      & LOSS       & $\surd$ & $\surd$ & $\surd$ &         &         &         &        \\
1998de & NGC 0252            & 1998-07-23 &  12.02867 &  27.62458 &  71.9E   3.4N & 18.4 & Ia-91bg & LOSS       & $\surd$ & $\surd$ & $\surd$ & $\surd$ & $\surd$ & $\surd$ & $\surd$\\
1998dh & NGC 7541            & 1998-07-20 & 348.66794 &   4.53725 &  53.5W  10.4N & 16.8 & Ia      & LOSS       & $\surd$ & $\surd$ &         & $\surd$ & $\surd$ &         & $\surd$\\
1998dj & NGC 0788            & 1998-08-08 &  30.27883 &  -6.81800 &   7.1E   8.6S & 16.1 & Ia      & LOSS       & $\surd$ & $\surd$ & $\surd$ &         &         &         &        \\
1998dk & UGC 00139           & 1998-08-19 &   3.63400 &  -0.73636 &   5.4E   3.1N & 17.6 & Ia      & LOSS       & $\surd$ & $\surd$ & $\surd$ & $\surd$ & $\surd$ & $\surd$ & $\surd$\\
1998dl & NGC 1084            & 1998-08-20 &  41.50613 &  -7.57364 &  21.6E  11.6N & 16.0 & II      & LOSS       & $\surd$ & $\surd$ & $\surd$ &         &         &         &        \\
1998dm & UGCA 017            & 1998-08-22 &  21.55821 &  -6.10389 &  13.8W  37.0S & 16.8 & Ia      & LOSS       & $\surd$ & $\surd$ &         & $\surd$ & $\surd$ &         & $\surd$\\
1998dn & NGC 0337A           & 1998-08-19 &  15.36283 &  -7.61019 & 102.0W  79.0S & 15.8 & II      & BAOSS      & $\surd$ & $\surd$ & $\surd$ &         &         &         &        \\
1998dt & NGC 0945            & 1998-09-01 &  37.14883 & -10.54994 &  23.1W  39.5S & 17.7 & Ib      & LOSS       & $\surd$ & $\surd$ & $\surd$ & $\surd$ & $\surd$ & $\surd$ & $\surd$\\
1998eb & NGC 1961            & 1998-09-17 &  85.55009 &  69.37397 &  38.8E  17.0S & 17.8 & Ia      & LOSS       & $\surd$ & $\surd$ & $\surd$ &         &         &         &        \\
1998ec & UGC 03576           & 1998-09-26 & 103.27546 &  50.03947 &   8.7W  19.5N & 16.9 & Ia      & BAOSS      & $\surd$ & $\surd$ & $\surd$ &         &         &         &        \\
1998ef & UGC 00646           & 1998-10-18 &  15.86196 &  32.23678 &   6.1E   2.1S & 15.2 & Ia      & LOSS       & $\surd$ & $\surd$ & $\surd$ & $\surd$ & $\surd$ & $\surd$ & $\surd$\\
1998en & UGC 03645           & 1998-10-30 & 106.00021 &  50.67500 &  10.4W  18.3S & 18.4 & II      & LOSS       & $\surd$ & $\surd$ &         & $\surd$ & $\surd$ &         &        \\
1998es & NGC 0632            & 1998-11-13 &  24.32292 &   5.88064 &   0.4W  10.8N & 14.6 & Ia-91T  & LOSS       & $\surd$ & $\surd$ & $\surd$ & $\surd$ & $\surd$ & $\surd$ & $\surd$\\
1998fa & UGC 03513           & 1998-12-25 & 100.71462 &  41.42192 &   4.0W   3.8N & 18.2 & IIb     & LOSS       & $\surd$ & $\surd$ & $\surd$ & $\surd$ & $\surd$ & $\surd$ &        \\
1999A  & NGC 5874            & 1999-01-10 & 226.98033 &  54.76111 &  29.9E  30.1N & 18.3 & II      & LOSS       & $\surd$ & $\surd$ & $\surd$ &         &         &         &        \\
1999D  & IC 0694             & 1999-01-16 & 172.11826 &  58.56083 &  18.9W   4.8S & 15.6 & II      & BAOSS      & $\surd$ & $\surd$ & $\surd$ & $\surd$ & $\surd$ & $\surd$ & $\surd$\\
1999aa & NGC 2595            & 1999-02-11 & 126.92512 &  21.48744 &   1.4E  31.0N & 15.5 & Ia-91T  & Arbour     & $\surd$ & $\surd$ & $\surd$ & $\surd$ & $\surd$ & $\surd$ & $\surd$\\
1999ac & NGC 6063            & 1999-02-26 & 241.81255 &   7.97233 &  23.9E  29.8S & 15.2 & Ia-91T  & LOSS       & $\surd$ & $\surd$ & $\surd$ & $\surd$ & $\surd$ & $\surd$ & $\surd$\\
\hline
\hline
\end{tabular}

\medskip

$^{a}$Only the first 20 entries are shown; the rest of the data
are available electronically. See text for details concerning the
meanings of the columns.

$^{b}$The date of discovery.

$^{c}$The offset (in arcsec) from the host galaxy nucleus. E = east, W = west, N = north, S = south.

$^{d}$The magnitude at discovery. 

$^{e}$The membership in the various SN subsamples. full-nos = full-nosmall, full-opt = full-optimal,
season-nos = season-nosmall, season-opt = season-optimal. 

\end{minipage}
\end{table*}

\end{landscape}